\documentclass[useAMS,usenatbib]{mn2e}
\usepackage{graphicx}
\usepackage{epsfig}
\usepackage{color}
\usepackage{tabularx}
\usepackage{multirow}
\usepackage[para,online,flushleft]{threeparttable}

\title[X-ray afterglows of GeV-TeV GRBs]{
Less noticeable shallow decay phase in early X-ray afterglows of
GeV/TeV-detected gamma-ray bursts
}
\author[Yamazaki et al.]
{Ryo~Yamazaki,$^{1,2}$\thanks{E-mail: ryo@phys.aoyama.ac.jp (RY)}
Yuri~Sato$^{1}$,
Takanori~Sakamoto$^{1}$,
Motoko~Serino$^{1}$
\\
$^{1}$Department of Physics and Mathematics, Aoyama Gakuin University, 5-10-1 Fuchinobe, Sagamihara 252-5258, Japan\\
$^{2}$Institute of Laser Engineering, Osaka University, 2-6 Yamadaoka, Suita, 565-0871 Osaka, Japan
}
\begin{document}
\date{}
\pagerange{\pageref{firstpage}--\pageref{lastpage}} \pubyear{2018}
\maketitle
\label{firstpage}

%%%  ABSTRACT
\begin{abstract}

The nature of the shallow decay phase in the X-ray afterglow of the gamma-ray burst (GRB) is not yet clarified.
We analyze the data of early X-ray afterglows of 26 GRBs triggered by Burst Alert Telescope onboard
{\it Neil Gehrels Swift Observatory} and
subsequently detected by {\it Fermi} Large Area Telescope (LAT) and/or
Imaging Atmospheric Cherenkov Telescopes.
It is found that 9 events (including 2 out of 3 very-high-energy gamma-ray events) have no shallow decay phase 
and that their X-ray afterglow
light curves are well described by single power-law model except for the jet break at later epoch.
The rest  are fitted by double power-law model and have a break in the early epoch (around ks), 
however, 8 events (including a very-high-energy gamma-ray event) have the pre-break decay index larger than 0.7. 
We also analyze the data of well-sampled X-ray afterglows of GRBs without LAT detection, and compare
their decay properties with those of high-energy and very-high-energy gamma-ray events.
It is found that for the GeV/TeV bursts, the fraction of events whose X-ray afterglows are described by single power-law
is significantly larger than those for non GeV/TeV GRBs.
Even if the GeV/TeV GRBs have shallow decay phase, their decay slope tends to be steeper than non GeV/TeV bursts, 
that is, they have less noticeable shallow decay phase in the early X-ray afterglow.
A possible interpretation along with the  energy injection model is briefly discussed.

\end{abstract}

%%%  KEYWORDS

%%%%%%%%%%%%%%%%%%%%%%%%%%%%%%%%%%%%%%%%%%%%%%%%%%%%%%%%%%%%%%%%%%%%%%%%%

%%%  INTRODUCTION
\section{Introduction}

%%%%%%%% RED COLOR %%%%%%%%%%%%%%

%\textcolor{red}{
%Test: red color
%}

%%%%%%%%%%%%%%%%%%%%%%%%%%%%%

X-ray afterglows of gamma-ray bursts (GRBs) are not fully understood as of yet
\citep[see, e.g.,][for review]{kumar2015}.
Their canonical behavior consists of the initial steep decay phase, 
the shallow decay phase and the normal decay phase
\citep{nousek2006,zhang2006}, which is subsequently followed  by the steepening again due to the jet break 
\citep{liang2008,racusin2009}.
The initial steep decay phase is most likely the tail emission of the prompt GRB
\citep{kumar2000,zhang2006,yamazaki2006}, and
the late normal decay phase is well explained by the external forward shock model
proposed in the pre-{\it Swift} era \citep{sari1998}.
The most enigmatic is the shallow decay phase which typically lasts $10^{3-4}$~s 
\citep{willingale2007,liang2007,sakamoto2008,dainotti2010,dainotti2013,dainotti2016,margutti2013,tang2019,zhao2019}.
Proposed models are
the energy injection model \citep{nousek2006,zhang2006,granot2006b,kobayashi2007},
the inhomogeneous or two-component jet model \citep{toma2006,eichler2006,granot2006,beniamini2019},
the time-dependent microphysics model \citep{ioka2006,granot2006,fan2006},
the prior explosion model \citep{ioka2006,yamazaki2009},
the cannonball model \citep{dado2006}, 
the reverse-shock-dominated afterglow model \citep{genet2007}, 
the internal engine model \citep{ghisellini2007},  
the supercritical pile model \citep{sultana2013},
the collapsar model ejecting thick shells \citep{vaneerten2014},
the high-latitude emission from structured jets \citep{oganesyan2019} 
and so on.
To clarify the mechanism of the shallow decay phase, additional observational information other than X-rays
is necessary.

High-energy gamma-ray emissions are detected by {\it Fermi} Large Area Telescope (LAT) either during or after
the prompt GRB emission, origin of which
 is still under debate \citep[see][for review]{nava2018}.
The early emission detected in the prompt phase may have an internal origin, and comes from
leptonic inverse-Compton process with various seed photons 
\citep{bosnjak2009,zhang2011,toma2011,asano2012,daigne2012,oganesyan2017}
or from hadronic process
\citep{asano2009,asano2012,razzaque2010}.
The temporally extended emission is likely the afterglow synchrotron emission
 arising in the external shock 
 \citep{kumar2009,kumar2010,ghisellini2010,nava2014}.
 Recently,  
 Imaging Atmospheric Cherenkov Telescopes (IACTs),
the Major Atmospheric Gamma Imaging Cherenkov (MAGIC) telescopes
and the High Energy Stereoscopic System (H.E.S.S.), 
detected very-high-energy (VHE) gamma-rays from
GRB~180720B \citep{abdalla2019},
GRB~190114C \citep{acciari2019}
and 
GRB~190829A \citep{denaurois2019}.
The VHE gamma-rays are most likely synchrotron self-Compton emission 
\citep{wang2019,derishev2019,fraija2019,MAGIC2019}. 
 It is expected that in near future the number of such {\it VHE events} rapidly increases when
 Cherenkov Telescope Array \citep[CTA;][]{acharya2013}
 starts observations of GRBs \citep{kakuwa2012,inoue2013,gilmore2013}. 
VHE gamma-ray observations with CTA may even provide a clue to the origin of shallow decay phase in the X-ray afterglow
\citep[e.g.,][]{murase2010,murase2011}.

At present, a link between the X-ray shallow decay phase and (very-)high-energy gamma-ray emission is unclear
\citep[e.g.,][]{panaitescu2008}.
It is known that the LAT-detected GRBs are among the most energetic GRBs 
\citep{racusin2011,ackermann2013,atteia2017,nava2014,ajello2019}, so that their kinetic energy of the GRB jet
may be larger than usual events without high-energy gamma-ray detection.
There is also an observational implication that the initial bulk Lorentz factor of the jet
is larger for LAT-detected GRBs \citep{ghirlanda2012}.
Furthermore, emission region of the high-energy gamma-rays might have smaller magnetic-field energy density
\citep{tak2019}.
Hence one can expect that their outflow 
has different characteristics, so that the X-ray afterglow behaves differently.
Therefore, studies of the X-ray afterglow of such extreme GRBs may provide us hints for unveiling the nature
of the shallow decay phase.

In this paper, as a first step of investigating connection between the shallow decay phase
and the high-energy (and VHE) gamma-ray emission, we analyze
early X-ray afterglows of GRBs with detected high-energy and VHE gamma-rays.
We find that their decay slopes of the shallow decay phase tend to be steeper than GRBs
without high-energy/VHE gamma-ray detection, so that the X-ray shallow decay phase looks less noticeable.
This fact has been already noted in previous literature very briefly \citep{kumar2015}.
Present work provides analysis result more quantitatively with better statistics due to larger sample size.

%%%%%%%%%%%%%%%%%%%%%%%%%%%%%%%%%%%%%%%%%%%%%%%%%%%%%%%%%%%%%%%%%%%%%%%%%

%%%  OBSERVATION AND DATA REDUCTION
\section{Sample Selection}

In this paper, we analyze early X-ray afterglows of 
GRBs that are detected by {\it Fermi}/LAT and VHE events (Sample~A).
For comparison, we also analyze the data of events without (very-)high-energy gamma-ray emission 
(Samples~B and C).
In the following, we define our samples.

\subsection{Sample~A: GeV/TeV events}

Candidate events in Sample~A are taken from
the second catalog of LAT-detected GRBs \citep{ajello2019}.
The catalog includes 186 events covering from 2008 to 2018 August~4. 
There are 25 events in the catalog
which were triggered by Burst Alert Telescope (BAT) onboard {\it Swift} and subsequently observed by X-ray Telescope (XRT) 
typically $\sim100$~s  after the burst onset.
Among them, XRT data of GRB~170813A consist of 
only 4 data points after the initial steep decay phase, so that we remove this event
from Sample~A in order to consider well-sampled early X-ray afterglow light curves.
Note that a VHE event GRB~180720B is also listed in  the catalog of \citet{ajello2019}.
Therefore, there are 23 events which were detected by {\it Fermi}/LAT but not detected by IACTs.

\begin{figure}
%\centering \vspace*{1pt}
\includegraphics[width=0.5\textwidth]{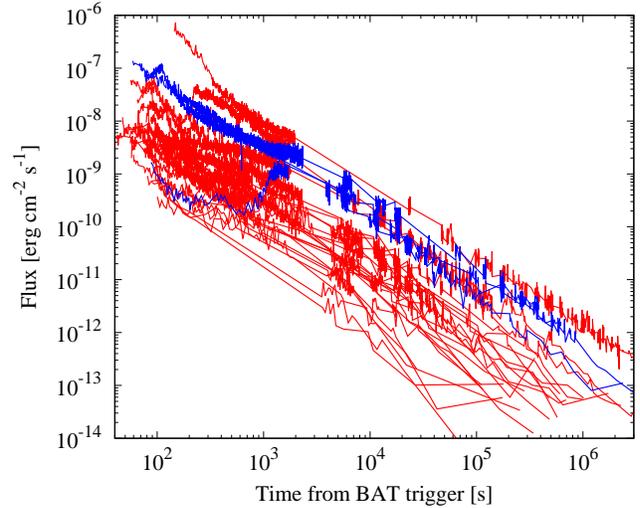}
\caption{
X-ray afterglow light curves of 26 events in 
Sample~A,
which consists of 23 events detected by {\it Fermi}/LAT but not detected by IACTs
(red curves) and 3~VHE events (blue lines). 
}
\label{fig:lc}
\end{figure}

So far VHE gamma-rays from 
3~GRBs (GRB~180720B, 190114C, and 190829A) are detected by MAGIC and 
H.E.S.S. 
\citep{abdalla2019,acciari2019,denaurois2019}.
Fortunately, all are triggered by {\it Swift}/BAT, so that early X-ray afterglows
are observed by XRT.
Hence, 
we include the three VHE events into Sample~A, and
 analyze XRT data of these events.
Note that
two of these (GRB~180720B and 190114C) were also detected by {\it Fermi}/LAT, while
for GRB~190829A, possible detections of sub-GeV photons have been claimed \citep{chand2020}.

All events of Sample~A 
are listed in Table~\ref{table1}, and their X-ray light curves are shown in Fig.~\ref{fig:lc}.
It contains 26 GRBs in total. 
%%(23 LAT GRBs $+$ 3 VHE events) 
%
\citet{tang2019} and \citet{zhao2019} collected 174 and 201 GRBs with clear shallow decay phase, respectively.
Within our Sample~A, only 4 events (GRB~090510, 110213A, 150403A and 180720B) 
overlap with the list of the former, and
3 events (GRB~090102, 090510 and 140323A) overlap with the latter.
This fact already shows that there are less events hosting typical shallow decay phase
in Sample~A.

Finally, we note that Sample~A contains a short GRB~090510.
The other events are long GRBs.

\begin{table*}
 \begin{minipage}{170mm}
%\centering
\caption{Best-fitting model parameters of GRBs in Sample~A.
}
\begin{tabular}{lcccccccc}
\hline
\hline
GRB  &  $t_1$--$t_2$ [ks]   &   \multicolumn{2}{c}{SPL model : $f_{\rm S}(t)$}
    &  \multicolumn{4}{c}{DPL model: $f_{\rm D}(t)$} &   $\Delta\chi^2$~${}^{\rm a}$ \\
%\hline
 &  & $\alpha_1$ & $\chi^2$/dof & $\alpha _1$ & $\alpha _2$ & $t_b$ [ks] & $\chi^2$/dof &  \\
\hline
%&&&&&&&& \\
\multicolumn{9}{l}{\bf Single power-law (SPL) events}  \\
%&&&&&&&& \\
081203A & 0.1--30 & $1.31\pm{0.02}$ & 1590/297  &  &  & & & $<1$ \\
110625A & 0.1--20 & $1.14\pm{0.03}$ & 90/53 &  & & & & $<1$ \\
121011A & 0.08--12 & $1.48\pm{0.03}$ & 38/19 &  & & & & $<1$ \\
151006A & 0.1--100 & $1.40\pm{0.01}$ & 159/157 & & & & & $<1$ \\
170405A & 0.2--10 & $1.40\pm{0.03}$ & 818/206 & & & & & $<1$ \\
\hline
%&&&&&&&& \\
\multicolumn{9}{l}{\bf Double power-law (DPL) events}  \\
%&&&&&&&& \\
090102 & 0.3--45 &$1.29\pm{0.01}$ & 143/121 & $0.21\pm{0.36}$ & $1.37\pm{0.03}$ & $0.62\pm{0.11}$ & 111/119 & 32 \\
090510${}^{\rm c}$ & 0.07--20 & $1.07\pm{0.05}$ & 460/70 & $0.64\pm{0.05}$ & $2.11\pm{0.12}$ & $1.46\pm{0.23}$ & 103/68 & 357 \\
100728A & 0.8--1000  & $1.26\pm{0.01}$ & 621/295 & $1.14\pm{0.04}$ & $1.65\pm{0.06}$ & $16.5\pm{6.88}$ & 499/293  & 122 \\
110213A & 0.15--1000 & $1.23\pm{0.03}$ & 2069/232 & $0.04\pm{0.06}$ & $1.82\pm{0.03}$ & $3.26\pm{0.17}$ & 400/230 & 1669 \\
110731A & 0.09--100 & $1.16\pm{0.01}$ & 346/268 & $1.13\pm{0.02}$ & $1.72\pm{0.22}$ & $25.1\pm{8.37}$ & 332/266 & 15 \\
120729A${}^{\rm b}$ & 0.05--40 & $1.18\pm{0.02}$ & 329/113 & $1.11\pm{0.02}$ & $2.82\pm{0.38}$ & $8.03\pm{1.38}$ & 202/111 & 128 \\
130427A & 0.35--1000 & $1.277\pm{0.003}$ & 2374/1409 & $1.18\pm{0.06}$ & $1.34\pm{0.04}$ & $3.97\pm{10.9}$ & 2335/1407 & 39 \\
130907A & 0.22--100 & $1.456\pm{0.004}$ & 4125/2296 & $1.36\pm{0.03}$ & $1.60\pm{0.05}$ & $5.80\pm{6.33}$ & 3937/2294 & 189 \\
140102A & 0.04--10 & $1.15\pm{0.01}$ & 804/524 & $1.05\pm{0.02}$ & $1.55\pm{0.07}$ & $1.71\pm{0.51}$ & 668/522 & 137 \\
140323A & 0.2--100 & $0.82 \pm{0.02}$ & 551/112 & $0.60\pm{0.03}$ & $1.67\pm{0.10}$ & $10.2\pm{1.76}$ & 184/110 & 367 \\
%% 150314A & 0.09--11 & $1.03\pm{0.01}$ & 1015/641 & $0.90\pm{0.01}$ & $1.47\pm{0.03}$ & $2.32\pm{0.36}$ & 951/672 & 65 \\
150314A & 0.09--110 & $1.08\pm{0.01}$ & 1532/674 & $0.90\pm{0.01}$ & $1.47\pm{0.03}$ & $2.32\pm{0.36}$ & 951/672 & 581 \\
150403A & 0.2--100 & $1.04\pm{0.01}$ & 6385/1601 & $0.43\pm{0.02}$ & $1.27\pm{0.01}$ & $1.28\pm{0.07}$ & 1813/1599 & 4572 \\
160325A & 0.2--10 & $1.36\pm{0.02}$ & 202/128 & $0.98\pm{0.29}$ & $1.53\pm{0.11}$ & $0.45\pm{0.47}$ & 182/126 & 20 \\
160905A & 0.1--100 & $0.96\pm{0.01}$ & 3796/989 & $0.66\pm{0.02}$ & $1.35\pm{0.02}$ & $1.48\pm{0.14}$ & 1495/987 & 2301 \\
160917A${}^{\rm b}$ & 0.07--25 & $1.25\pm{0.02}$ & 52.4/32 & $1.22\pm{0.03}$ & $2.33\pm{0.72}$ & $12.2\pm{4.25}$ & 42.2/30 & 10.2 \\
170728B & 0.3--20 & $0.99\pm{0.01}$ & 374/196 & $0.32\pm{0.17}$ & $1.22\pm{0.03}$ & $1.21\pm{0.32}$ & 275/194 & 99 \\
170906A & 0.2--30 & $1.26\pm{0.03}$ & 2055/267 & $0.35\pm{0.05}$ & $1.91\pm{0.14}$ & $1.14\pm{0.30}$ & 450/265 & 1605 \\
171120A & 3--150 & $0.61\pm{0.04}$ & 173/75 & $0.39\pm{0.06}$ & $1.77\pm{0.24}$ & $37.0\pm{8.2}$ & 87/73 & 85 \\
\hline
%&&&&&&&& \\
\multicolumn{9}{l}{\bf Very-high-energy gamma-ray (VHE) events}  \\
%&&&&&&&& \\
180720B & 0.25--100 & $0.931\pm{0.004}$ & 7233/3361 & $0.74\pm{0.01}$ & $1.43\pm{0.02}$ & $4.70\pm{0.23}$ & 4731/3359 & 2502 \\
190114C & 0.065--100 & $1.338\pm{0.004}$ & 1822/1030 & & & & & $<1$ \\
190829A & 2--100 & $1.33\pm{0.02}$ & 451/235 & & & & & $<1$ \\
\hline
\end{tabular}
\begin{tablenotes}
\item {\bf Notes.} \\
${}^{\rm a}$~A DPL model is statistically preferred at $>3\sigma$ over a simpler SPL model when $\Delta\chi^2>10$.\\
${}^{\rm b}$~Best-fitting values of $\alpha_1$ and $\alpha_2$ of DPL model are consistent with the jet break (see section~5).\\
${}^{\rm c}$~A short GRB.\\
\end{tablenotes}
\label{table1}
\end{minipage}
\end{table*}

\subsection{Sample~B: non-GeV events~1}

We select GRBs that were triggered by {\it Swift}/BAT during ten years from 2008 to 2018 August~4,
followed-up by XRT within 400~s after the trigger,
but not detected by {\it Fermi}/LAT.
Among them, 180 long and 3 short bursts have measured redshifts.
%% and isotropic-equivalent gamma-ray energy in 15--150~keV.
We remove 13 events of which the number of X-ray data points after the initial steep decay phase is less than ten.
We also discard 15 GRBs showing the X-ray flux rising more steeply than $\propto t^1$ as being of the different origin.
Then, 155 events are remaining, and they are defined as members of Sample~B and are listed in Table~2.

Although the size of Sample~B is large, its members are 
{\it either} bursts that were not within the LAT field of view
{\it or} bursts that LAT did observe the GRB field but did not detect high-energy photons.
It is also noted that for Sample~A the redshift measurement has not been introduced, while for Sample~B
it has.
We believe that the latter fact does not bias our conclusion because the redshift can be determined if the event is observable
with ground-based optical telescopes and this is irrelevant if the event is detected with {\it Swift}/BAT and/or {\it Fermi}/LAT.
However, one might still concern that it causes some bias on the energetics/brightness of  Samples~A and B.
Therefore, we consider another Sample~C as in the next subsection.

\subsection{Sample~C: non-GeV events~2}

\citet{ackermann2016} selected 79 GRBs, observed between 2008 August 4 and 2012 February 1,
which fell in the LAT field of view at the time of trigger, but were not detected by LAT.
There are 36 events in their sample whose XRT data are well, satisfying the same criteria as Samples~A and B.
These bursts are defined as members of Sample~C, and listed in Table~\ref{table3}.
Sample~C contains no short GRBs.

The observation period during which Sample~C bursts were detected was shorter than that for Sample~B.
There are 67 Sample~B events detected in the same time period for Sample~C (from GRB~080804 to 120119A),
and among them only 13 GRBs are in the list of \citet{ackermann2016}.
Hence, $13/67\approx 19$~\% of Sample~B events were in the LAT field of view at the time of trigger but they were not detected by LAT.

%%%%%%%%%%%%%%%%%%%%%%%%%%%%%%%%%%%%%%%%%%%%%%%%%%%%%%%%%%%%%%%%%%%%%%%%%

\section{Data Analysis}

The {\it Swift}/XRT data were downloaded from the {\it Swift} team 
website\footnote{https://www.swift.ac.uk/xrt\_curves/} 
\citep{evans2007,evans2009}.
First the X-ray light curves in the time interval $[t_1,t_2]$ are fitted with 
single power-law (SPL) function, 
\begin{equation}
f_{\rm S} (t)= f_0 t^{-\alpha_1}~~,
\label{eq:SPL}
\end{equation}
where $f_0$ and $\alpha_1$ are a normalization constant and a decay slope,
respectively.
We choose the time interval $[t_1,t_2]$ excluding the steep decay phase and X-ray flares
if they exist.
Subsequently, we also fit the light curves with double power-law (DPL) function,
\begin{equation}
f_{\rm D} (t)= f_0 \left[ (t/t_{\rm b})^{w\alpha_1}+(t/t_{\rm b})^{w\alpha_2}\right]^{-1/w}  ~~,
\label{eq:DPL}
\end{equation}
with $\alpha_1$ and $\alpha_2$ describing the decay slopes of pre- and post-break segments, respectively,
and $t_{\rm b}$ is a break time.
A smoothness parameter $w$ is fixed to be 3 \citep{liang2007,zhao2019}.

We compare above two models in order to determine whether the additional degrees of freedom in the DPL model are
warranted over a simpler SPL model.
We fit all light curves of our events with both models and obtain
 $\chi^2$ of the best-fitting parameter set.
 Then, the difference between $\chi^2$ of the two models, $\Delta\chi^2$, is calculated.
Since there are two additional free parameters between the two models, 
a value of $\Delta\chi^2>10$ would represent a $>3\sigma$ improvement in the fit. 
We adopt this criterion as the threshold for a statistical
preference for a break in the light curve.

%%%%%%%%%%%%%%%%%%%%%%%%%%%%%%%%%%%%%%%%%%%%%%%%%%%%%%%%%%%%%%%%%%%%%%%%%

%\newpage

%%% RESULTS
\section{Results}

The results of the analysis on 26 events in Sample~A are shown in Table~\ref{table1}.
The upper most 5 events in the table are fitted with both SPL and DPL models,
however, we find $\Delta\chi^2<1$, so that additional two  parameters of the DPL model
do not improve the fit. 
We call them {\it SPL events}.
The other events in the table except for the lowest three VHE events have $\Delta\chi^2>10$, so that the DPL
model is statistically preferred at $>3\sigma$ over SPL model.
Hence, they are named as {\it DPL events} in this paper.
We also analyze 3 VHE events in a similar manner, and find that
two events (GRB~190114C and 190829A) are fitted with the SPL model and 
that the other one (GRB~180720B) is described by the DPL model.
The  average value of $\alpha_1$ for 18 DPL events is $\langle\alpha_1\rangle=0.76$ with
the standard deviation $\sigma_{\alpha_1}=0.40$.
When 5 SPL events are incorporated, the average value and the standard deviation 
become $\langle\alpha_1\rangle=0.89$ and $\sigma_{\alpha_1}=0.43$, respectively.
Further addition of 3 VHE events makes $\langle\alpha_1\rangle=0.92$
and $\sigma_{\alpha_1}=0.42$.

Similarly, the results for Sample~B are shown in Table~2.
Again we define SPL and DPL events as those having $\Delta\chi^2<10$ and $>10$, respectively.
It is found that 41 out of 155 events (26~\%) are fitted with SPL.
The  average value of $\alpha_1$ for 114 DPL events is $\langle\alpha_1\rangle=0.37$ and
their standard deviation is $\sigma_{\alpha_1}=0.43$.
When 41 SPL events are incorporated, the average value becomes $\langle\alpha_1\rangle=0.54$
with the standard deviation $\sigma_{\alpha_1}=0.51$.
We also show the results for Sample~C in Table~3.
Among 36 events, only 5 events ($5/36\approx14$~\%) are explained by SPL model.
The average value of $\alpha_1$ for 31 DPL events is
$\langle\alpha_1\rangle=0.28$ with $\sigma_{\alpha_1}=0.34$.
Incorporating the 5 SPL events, we get 
$\langle\alpha_1\rangle=0.40$ and $\sigma_{\alpha_1}=0.42$.

\begin{figure}
%\centering \vspace*{1pt}
%\includegraphics[width=0.5\textwidth]{fig2.eps}
\includegraphics[width=0.5\textwidth]{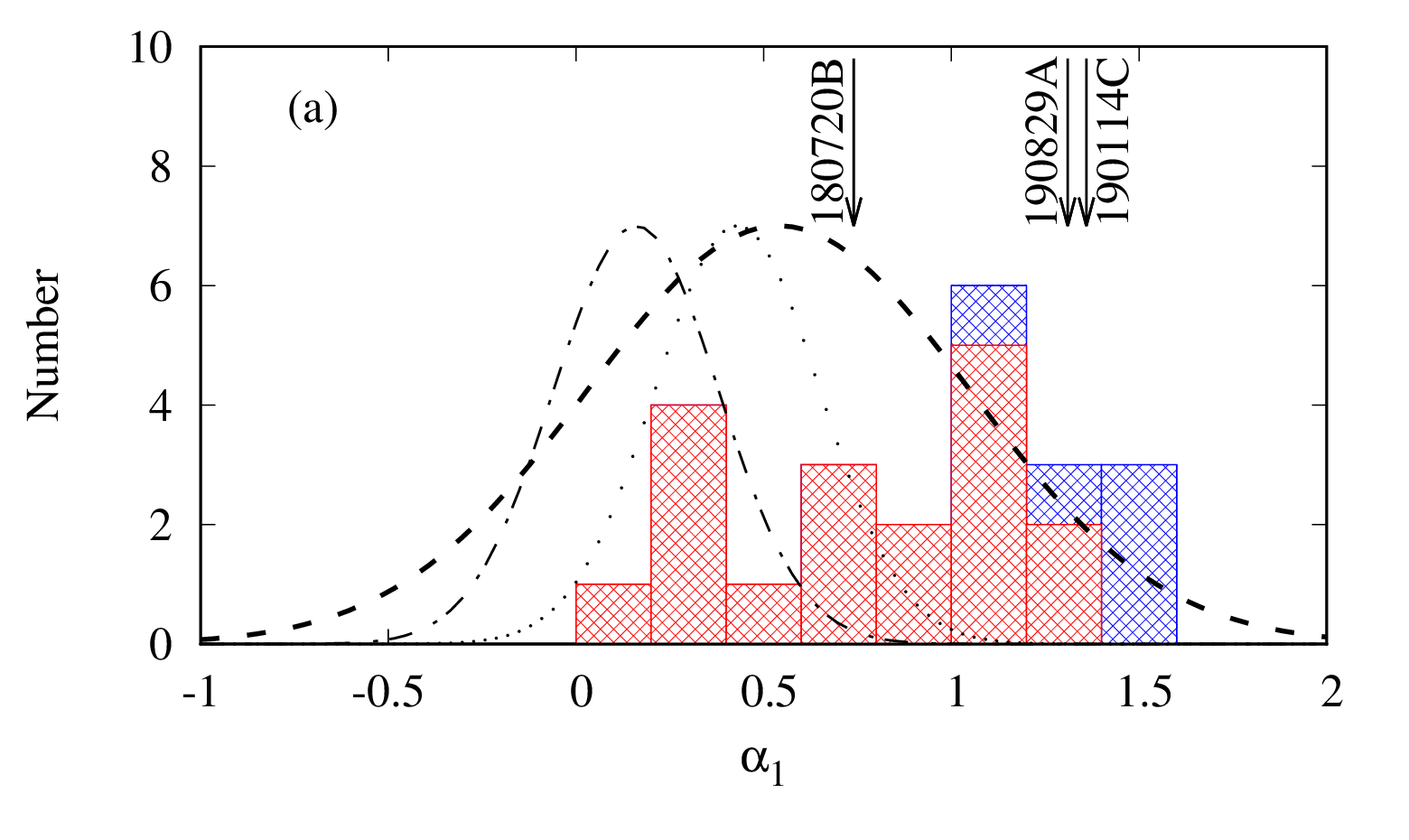}
\includegraphics[width=0.5\textwidth]{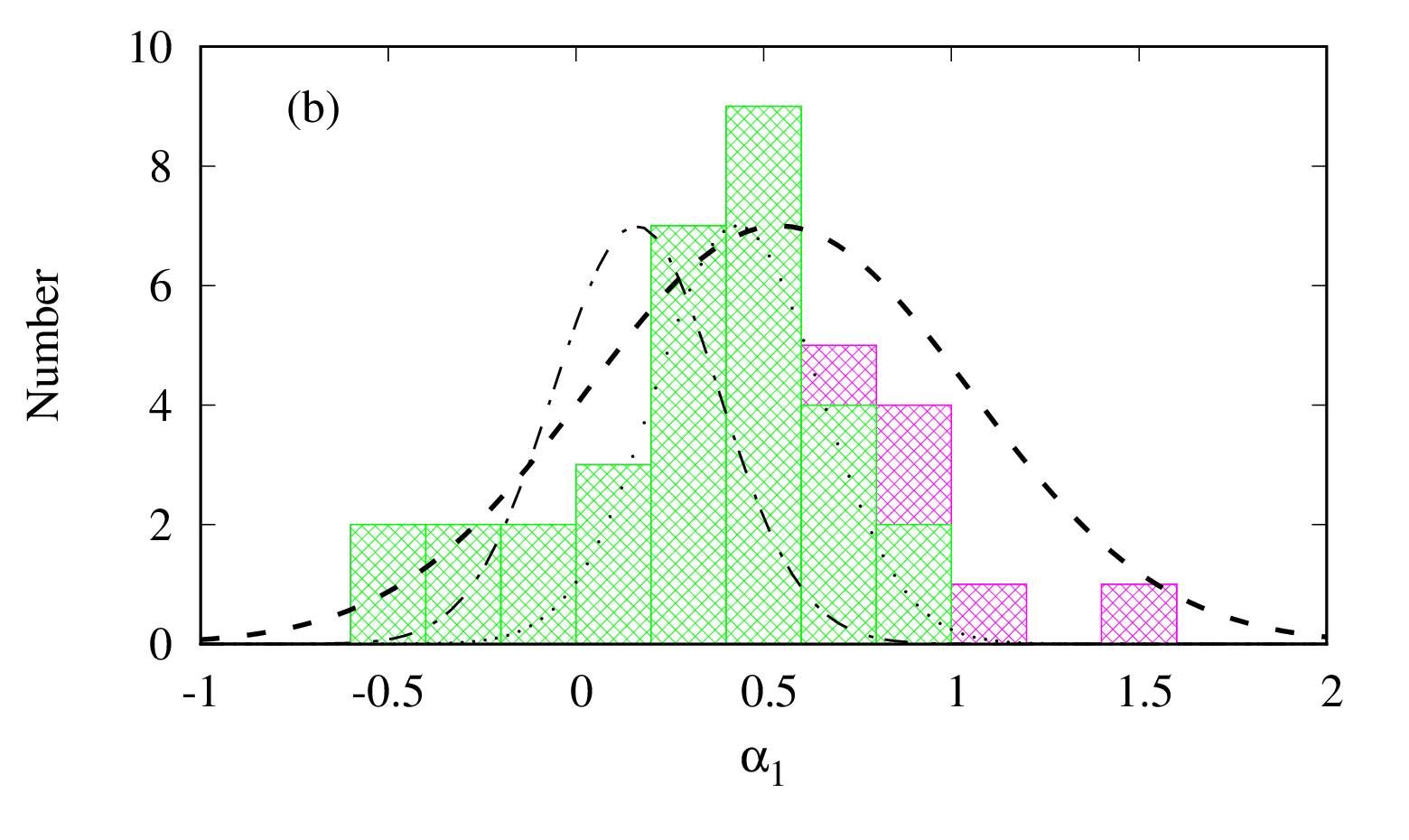}
\caption{
(a)
Blue and red histograms show
the distributions of the decay slope $\alpha_1$ for 5 SPL  
and 18 DPL events, respectively,
in Sample~A.
Thick-dashed line represents the same distribution for the whole Sample~B events
(155 events; $\langle\alpha_1\rangle=0.54$ and $\sigma_{\alpha_1}=0.51$).
Thin-dot-dashed and thin dotted lines 
are those for long GRBs with typical shallow decay phase 
taken from \citet{tang2019} and \citet{zhao2019}, respectively.
Also shown are arrows describing the values of $\alpha_1$ for 3 VHE events.
(b)~ Magenta and green histograms show the $\alpha_1$ distribution of 5~SPL and
31~DPL events, respectively, in Sample~C.
The three lines are the same as those in panel~(a).
}
\label{fig:alpha1dist}
\end{figure}

Histogram in Figure~\ref{fig:alpha1dist}(a)
shows the distribution of $\alpha_1$ for 5 SPL and 18 DPL events in Sample~A.
For the DPL events we take the best-fitting values of $\alpha_1$ of the DPL model rather than the SPL model.
Thick-dashed line represents that of Sample~B.
We find that 14 out of 26 events of Sample~A have $\alpha_1>1.05$, that is, they depart from more than $1\sigma$ of
the whole events of Sample~B.
The other two Gaussian distributions (thin-dot-dashed and thin-dotted lines) 
are taken from \citet{tang2019} and \citet{zhao2019},
describing the distribution of the temporal index of the typical shallow decay phase.
It is found that events in Sample~A
(including VHE events shown by arrows) tend to have larger value of $\alpha_1$
than those with typical shallow decay phase.
It is also noted that the SPL events sit the upper end of the $\alpha_1$ distribution.
If $\alpha_1>1$, the decay phase is no longer the shallow decay phase, but the normal decay phase.
The same plot but for Sample~C is shown in Figure~\ref{fig:alpha1dist}(b).
It is indicated that Sample~C events have smaller value of $\alpha_1$ than those of Samples~A and B.

\begin{figure*}
\begin{minipage}{0.45\linewidth}
\centering
\includegraphics[width=1.0\textwidth]{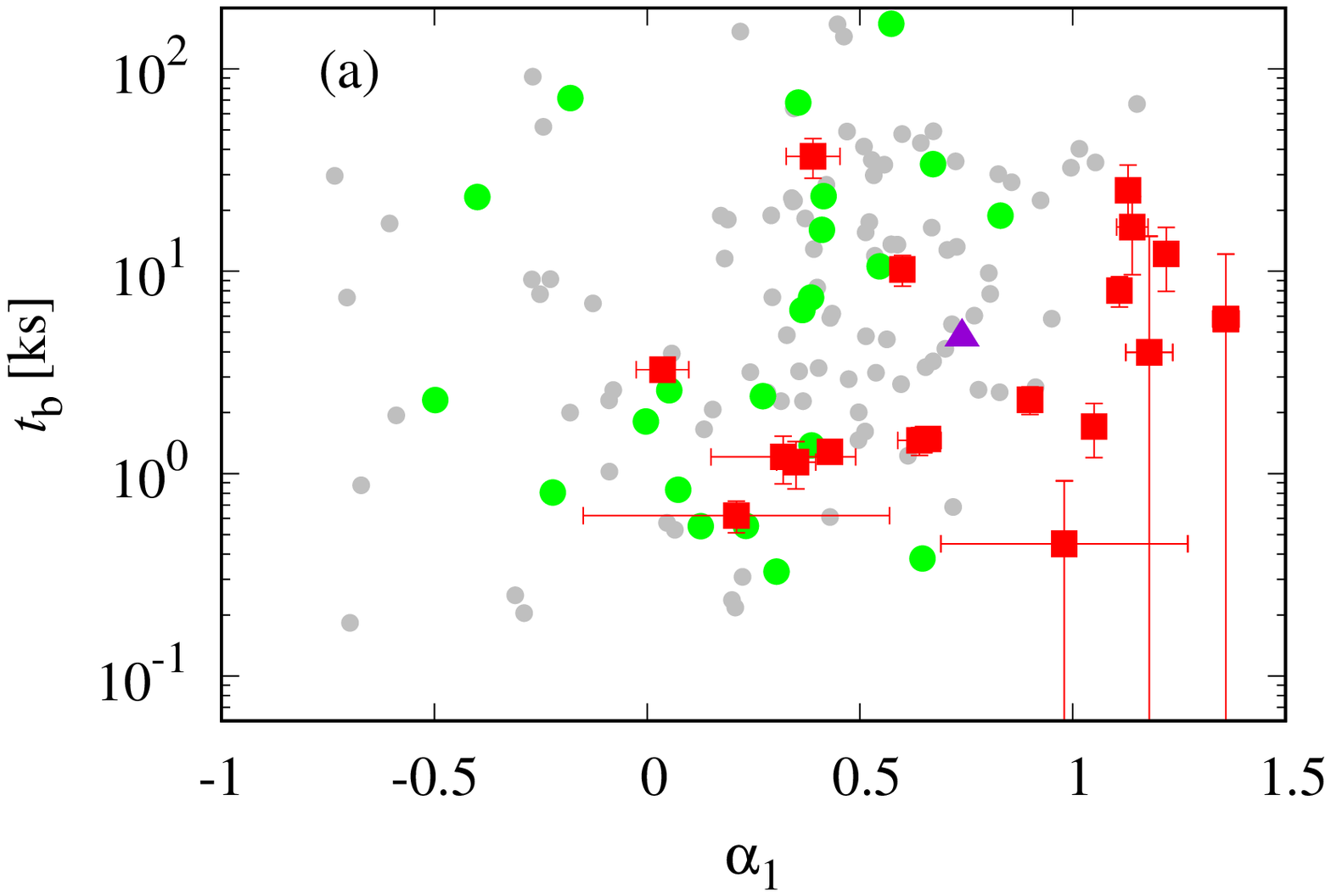}
\end{minipage}
\begin{minipage}{0.45\linewidth}
\centering
\includegraphics[width=1.0\textwidth]{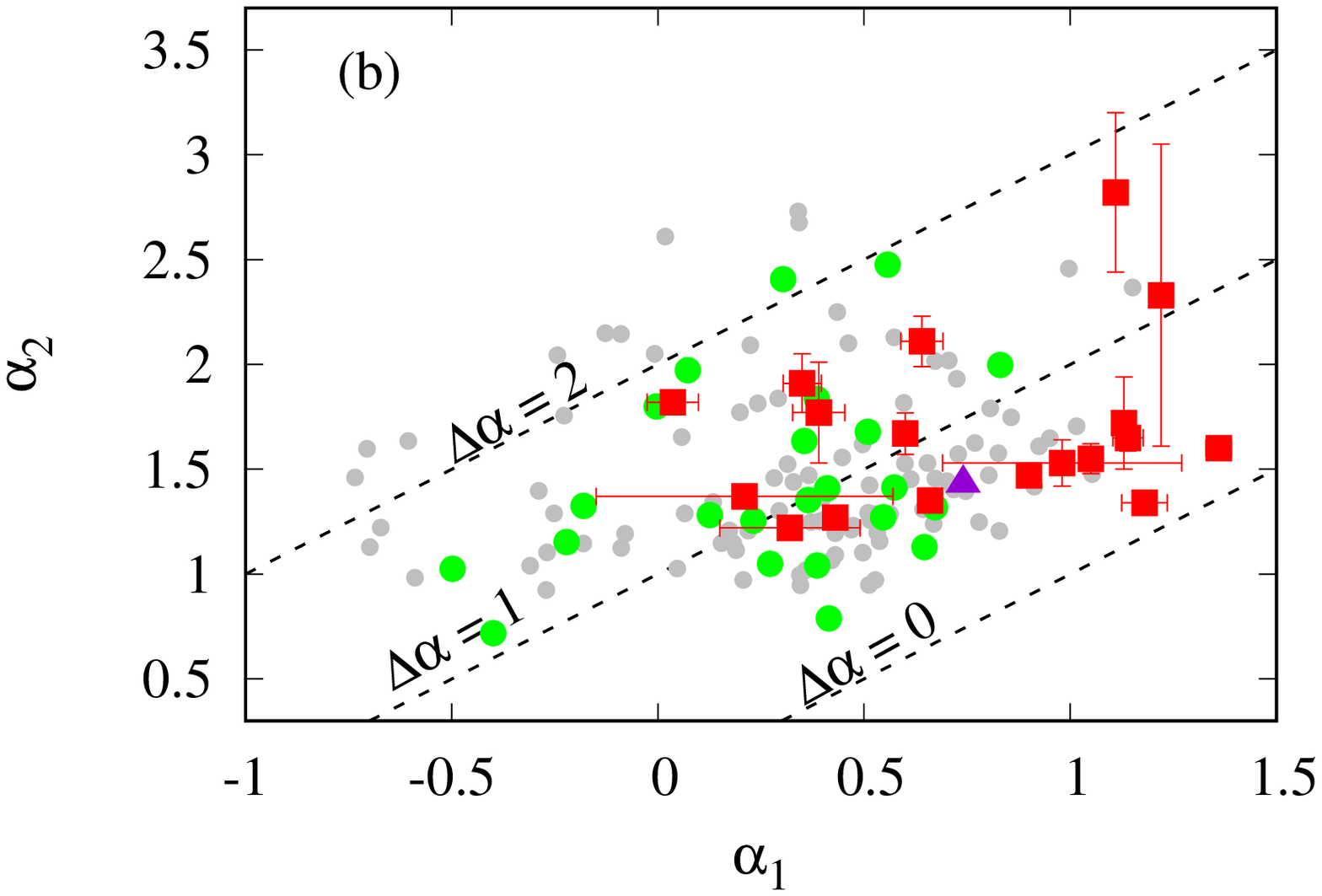}
\end{minipage}
\caption{
Comparison of 18 DPL events (red squares) of Sample~A and a VHE event (GRB~180720B: violet triangle) to bursts of 
Sample~B (smaller grey dots) and Sample~C (larger green dots)
in $\alpha_1$--$t_b$ plane (left panel) and 
$\alpha_1$--$\alpha_2$ plane (right panel).
}
\label{fig:DPLdist}
\end{figure*}

\begin{figure*}
\begin{minipage}{0.45\linewidth}
\centering
\includegraphics[width=1.0\textwidth]{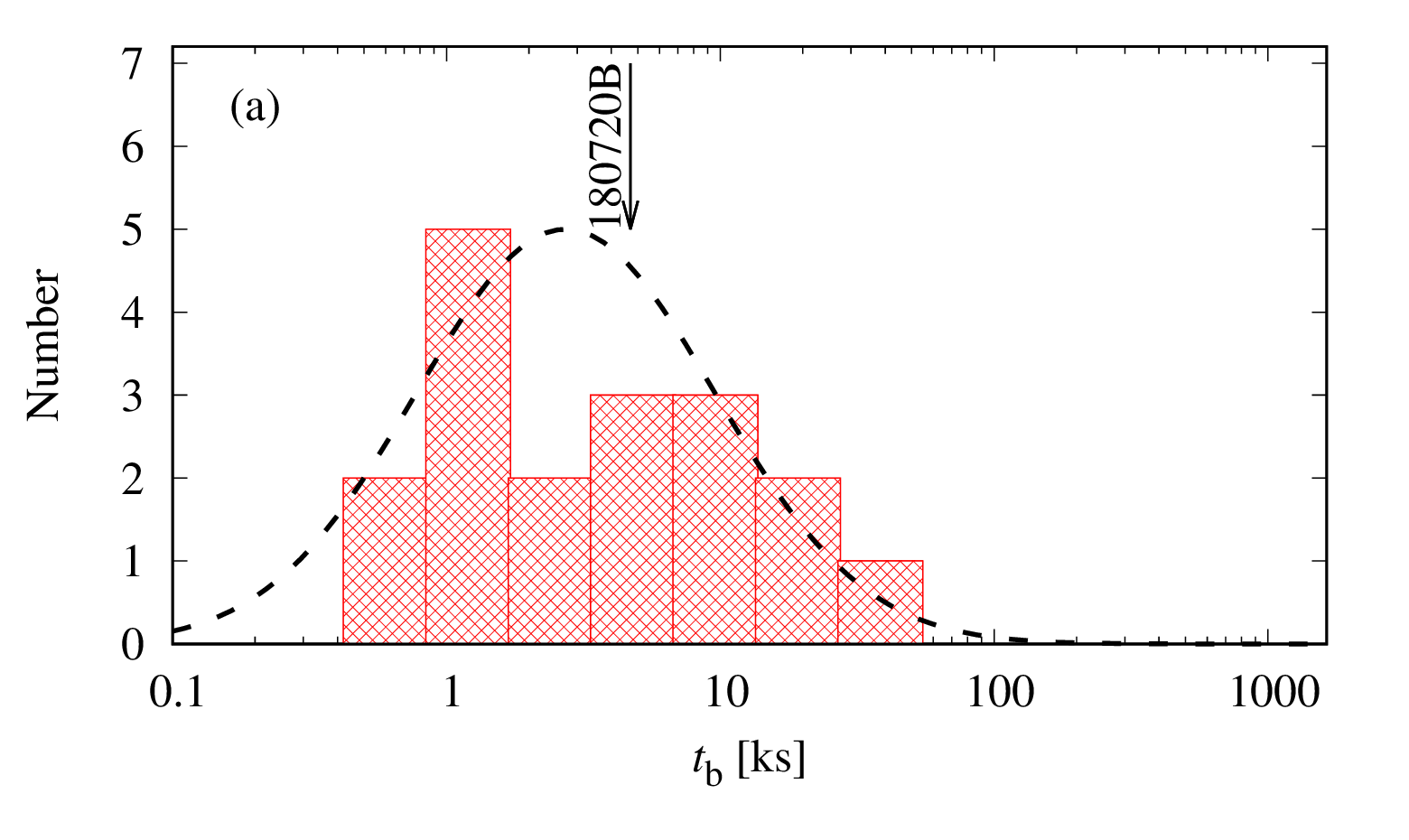}
\end{minipage}
\begin{minipage}{0.45\linewidth}
\centering
\includegraphics[width=1.0\textwidth]{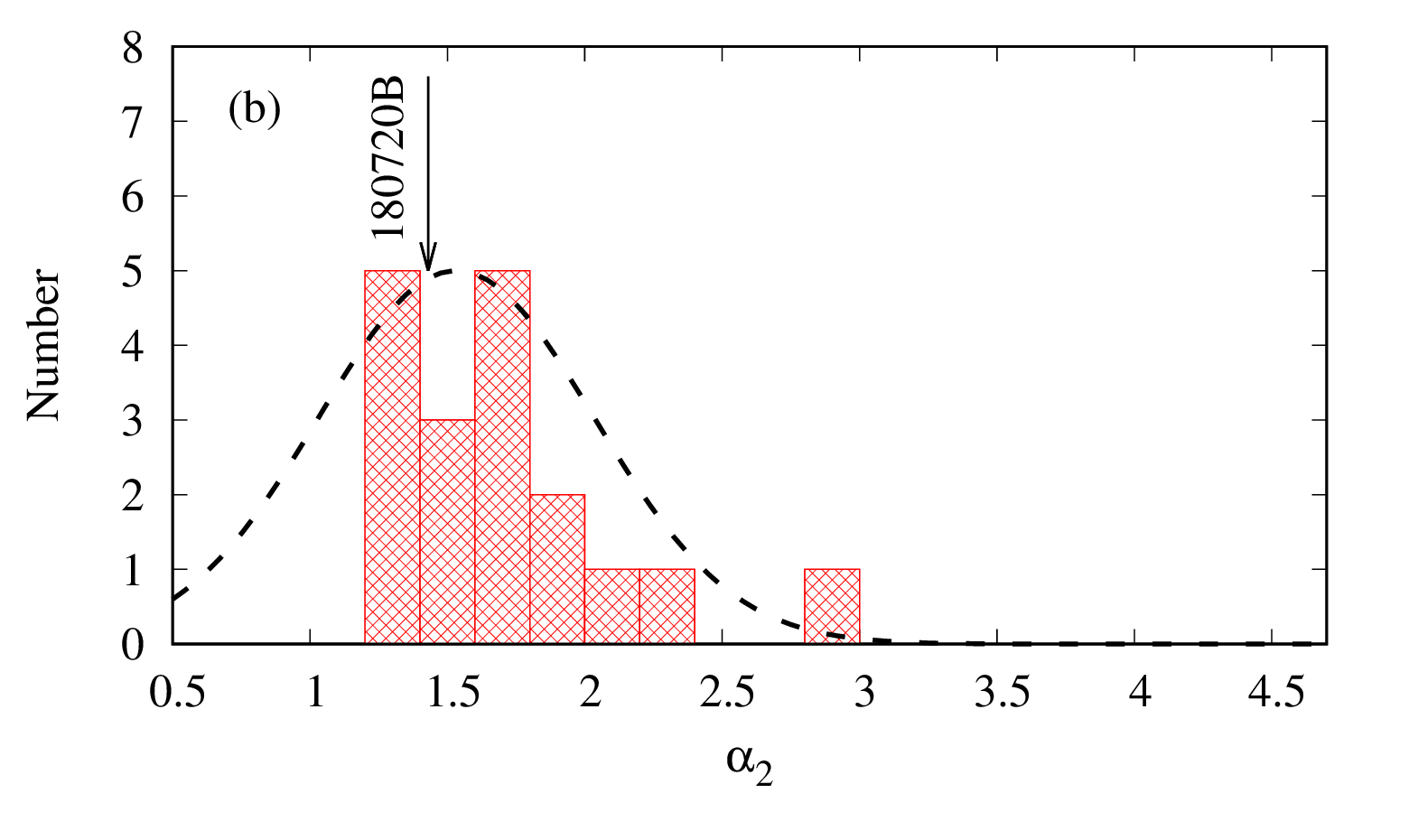}
\end{minipage}
\begin{minipage}{0.45\linewidth}
\centering
\includegraphics[width=1.0\textwidth]{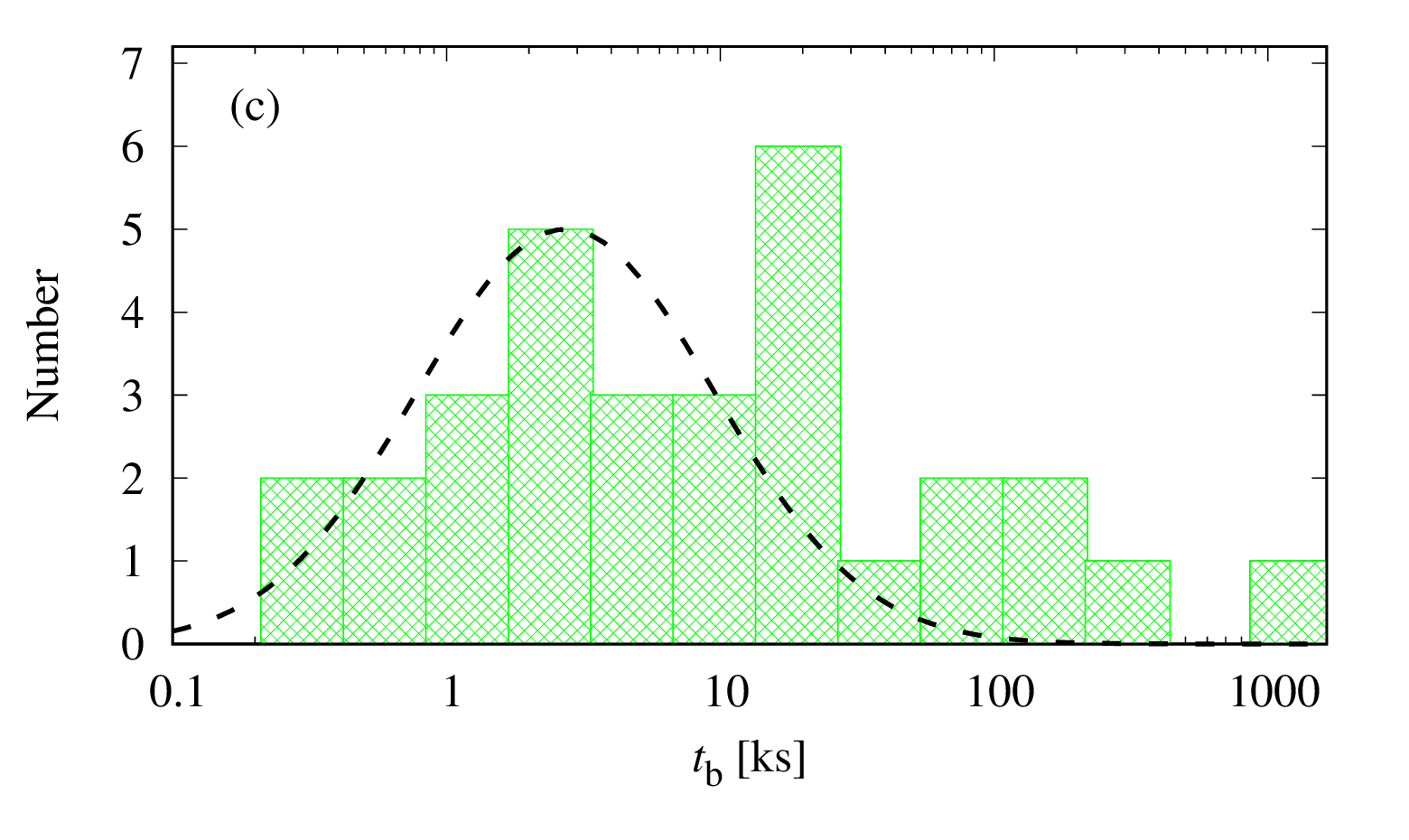}
\end{minipage}
\begin{minipage}{0.45\linewidth}
\centering
\includegraphics[width=1.0\textwidth]{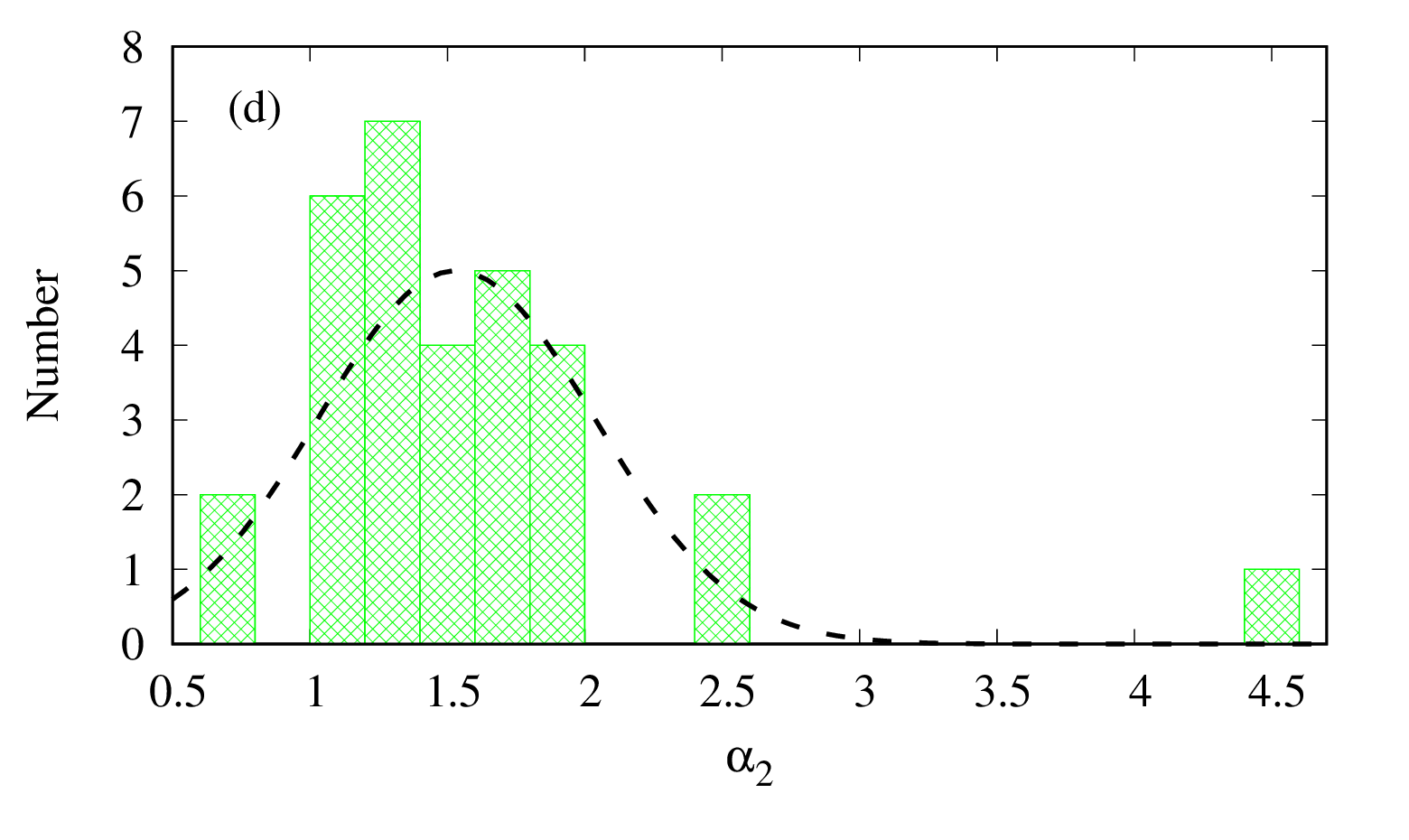}
\end{minipage}
\caption{
Red histograms in panels (a) and (b) show
the distributions of the break time $t_b$ and the post-break decay index $\alpha_2$
of 18 DPL events in Sample~A, respectively.
Panels (c) and (d) are the same as (a) and (b), respectively, but for Sample~C.
Thick-dashed lines are for 114 DPL events in Sample~B. 
For Sample~B, the mean and the standard deviation of $t_b$ are $\langle \log_{10} (t_b/{\rm ks})\rangle=0.44$ and $\sigma_{t_b}=0.54$~dex, respectively
[panels (a) and (c)],
and the mean and the standard deviation of $\alpha_2$ are $\langle\alpha_2\rangle=1.53$ and $\sigma_{\alpha_2}=0.50$, respectively
[panels (b) and (d)].
Also shown in panels (a) and (b) are arrows describing the values of a VHE event, GRB~180720B.
}
\label{fig:DPLdist_hist}
\end{figure*}

To see the decay properties of the DPL events in more detail,
%we make Figures~\ref{fig:alpha1tb} and \ref{fig:alpha1alpha2}, where
we show in Figure~\ref{fig:DPLdist}
the best-fitting parameters ($\alpha_1$, $\alpha_2$ and $t_{\rm b}$) of DPL events 
of Samples~A (red squares) 
in the  $\alpha_1$--$t_b$ (left panel) and  $\alpha_1$--$\alpha_2$ planes (right panel).
Smaller grey dots are those of Sample~B, while larger green dots are of Sample~C.
%% with clear shallow decay phase whose data are taken from  \citet{tang2019}.
It is found from Fig.~\ref{fig:DPLdist} that compared with events of Sample~B,
roughly a half of 18 Sample~A DPL events as well as a VHE event (GRB~180720B)
have larger pre-break decay index $\alpha_1$ while
the break time $t_{\rm b}$ and the post-break decay index $\alpha_2$ 
are roughly similar.
Figure~\ref{fig:DPLdist_hist} supports this claim.
More quantitatively,
15  out of 18 Sample~A DPL events as well as the VHE event GRB~180720B
have values of $\alpha_2$ within $1\sigma$ range of Sample~B (dashed line in the right panel of Figure~\ref{fig:DPLdist_hist}),
and 13 out of the 19 Sample~A events have values of $t_b$ within
$1\sigma$ range of Sample~B (dashed line in the left panel of Figure~\ref{fig:DPLdist_hist}).
According to these results, we schematically draw in Figure~5 %\ref{fig:lctyp} 
the typical behavior of GRBs in Sample~A.

%%%%%%%%%%%%%%%%%%%%%%%%%%%%%%%%%%%%%%%%%%%%%%%%%%%%%%%%%%%%%%%%%%%%%%%%%

%%%%%%%%%%%%%%%%%%%%%%%%%%%%%%%%%%%%%%%%%%%%%%%%%%%%%%%%%%%%%%%%%%%%%%%%%

%%%%%%%%%%%%%%%%%%%%%%%%%%%%%%%%%%%%%%%%%%%%%%%%%%%%%%%%%%%%%%%%%%%%%%%%%

\begin{figure}
\begin{center}
\includegraphics[width=0.35\textwidth, angle=270]{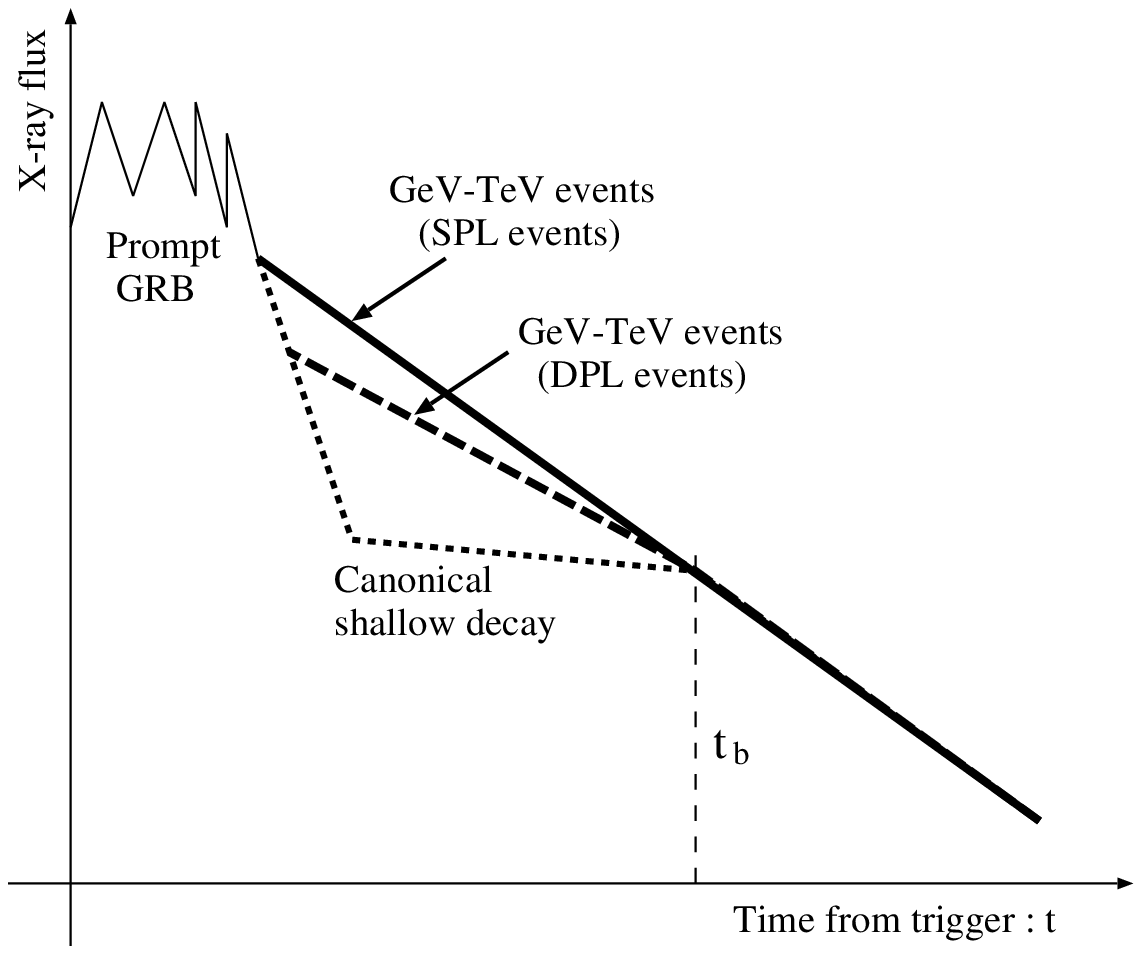}
\caption{
Schematic view of early X-ray afterglow light curves of  26 events considered in this paper.
Nine events (5~SPL and 2 DPL events as well as 2 VHE events) have no shallow decay phase.
Eight events (7 DPL events and a VHE event) have a break but their pre-break index $\alpha_1>0.7$, 
so that the decay slope before the break time $t_{\rm b}$
is somewhat steeper than typical shallow decay phase.
}
\end{center}
\label{fig:lctyp}
\end{figure}

%%%%%%%%%%%%%%%%%%%%%%%%%%%%%%%%%%%%%%%%%%%%%%%%%%%%%%%%%%%%%%%%%%%%%%%%%

%%% DISCUSSION
\section{Discussion}

In the $\alpha_1$--$\alpha_2$ plane for DPL events,
there are two data points from Sample~A 
(GRB~120729 and 160917A)  whose best-fitting values $\alpha_1>1$ and $\alpha_2>2$.
Although the break time $t_{\rm b}\sim10^4$~s for these bursts,
the measured break should be taken as a jet break rather than the {\it shallow-to-normal} break.
According to the theory of the jet break, if the X-ray afterglow is in the slow cooling regime with
the X-ray band frequency larger than the cooling frequency $\nu_{\rm c}$, then
the decay indices are given by $(3p-2)/4$ and $p$ for pre- and post-jet break, respectively,
where $p$ is an index of the power-law electron distribution \citep{sari1999}.
For GRB~160917A, if the measured value of $\alpha_1=1.22\pm0.03$ corresponds to the
pre-jet break decay index, then we have $p=(2+4\alpha_1)/3=2.29\pm0.04$, so that
the observed value of $\alpha_2=2.33\pm0.72$ is consistent with the post-jet break decay index within $1\sigma$ error.
On the other hand, GRB~120729A may not follow the jet break theory.
Similar calculation for GRB~120729A leads to $p=(2+4\alpha_1)/3=2.15\pm0.03$, 
which is somewhat smaller than
the measured post-break index $\alpha_2=2.82\pm0.38$ but still consistent within $2\sigma$ error.
Nevertheless, the post-break decay index $\alpha_2$ is too steep for the normal decay phase of the X-ray afterglow.
Therefore, these bursts should be treated as events without shallow decay phase.
Also in Sample~B, there are two events (GRB~110818A and 120119A) having $\alpha_1>1$ and $\alpha_2>2$.
They are also regarded as bursts without shallow decay phase.

Taking into account the correction described in the previous paragraph, we calculate for Sample~A
the fraction of events without shallow decay phase as
5~SPL as well as 2~DPL events (GRB~120729 and 160917A) out of 23 events, so that
$7/23\approx30\%$.
In order to compare this fraction with that of Sample~B, we need more careful treatment of 41 bursts
in Sample~B being claimed as SPL events.
Among them, there are 18 events which show only X-ray flares until 3~ks after the burst trigger, possibly overlying shallow and normal decay phases.
In addition, there are 4 bursts showing longer initial steep decay phase until 2~ks without steep-to-shallow break.
These 22 events should be excluded, and we regard the remaining 19 events as bursts without clear shallow decay phase.
Hence, the fraction of events without shallow decay phase as these 19 events as well as 2 DPL events described in the previous paragraph
(GRB~110818A and 120119A)
out of 155 events, so that $21/155\approx14\%$.
The same argument can be done for Sample~C.
Among 5~SPL events, there are 3 bursts showing only X-ray flares until 3~ks, and there is
an event with long initial steep decay phase lasting 2~ks (see Table~3).
Then, we get the fraction of events without shallow decay phase as $1/36\approx3$~\%.
Although sample selection method is different from ours,
\citet{liang2009} reported the number, $19/400\approx5\%$, for 
all long GRBs with XRT detection from 2005 January to 2009 July.

Furthermore, two (GRB~190114C and 190829A) out of 3~VHE events detected so far
have no shallow decay phase.
Even if X-ray light curve has a break, 12 events out of 18 in Sample~A
%%(GRB~100728A, 110731A, 130427A, 130907A, 140102A, 150314A, 160325A and a VHE event GRB~180720B)
have the pre-break decay index $\alpha_1$ larger than 
the mean value, $\langle\alpha_1\rangle=0.37$, of DPL events in Sample~B.
In addition,  \citet{zhao2019} derived a mean of $\langle\alpha_1\rangle=0.43$ and
a dispersion of 0.22 (see thin-dotted line in Fig.~\ref{fig:alpha1dist}) for events showing typical shallow decay phase.
If we take into account their result, 8 DPL events in our Sample~A have $\alpha_1>0.7$, 
deviating from the mean of \citet{zhao2019} more than 1$\sigma$.
Hence one can say  that a large fraction (17 out of 26 events) of
GRBs detected in high-energy and VHE gamma-ray bands has no shallow decay phase, or they have
less noticeable shallow decay phase in the early X-ray afterglow.

\begin{figure}
%\centering \vspace*{1pt}
\includegraphics[width=0.5\textwidth]{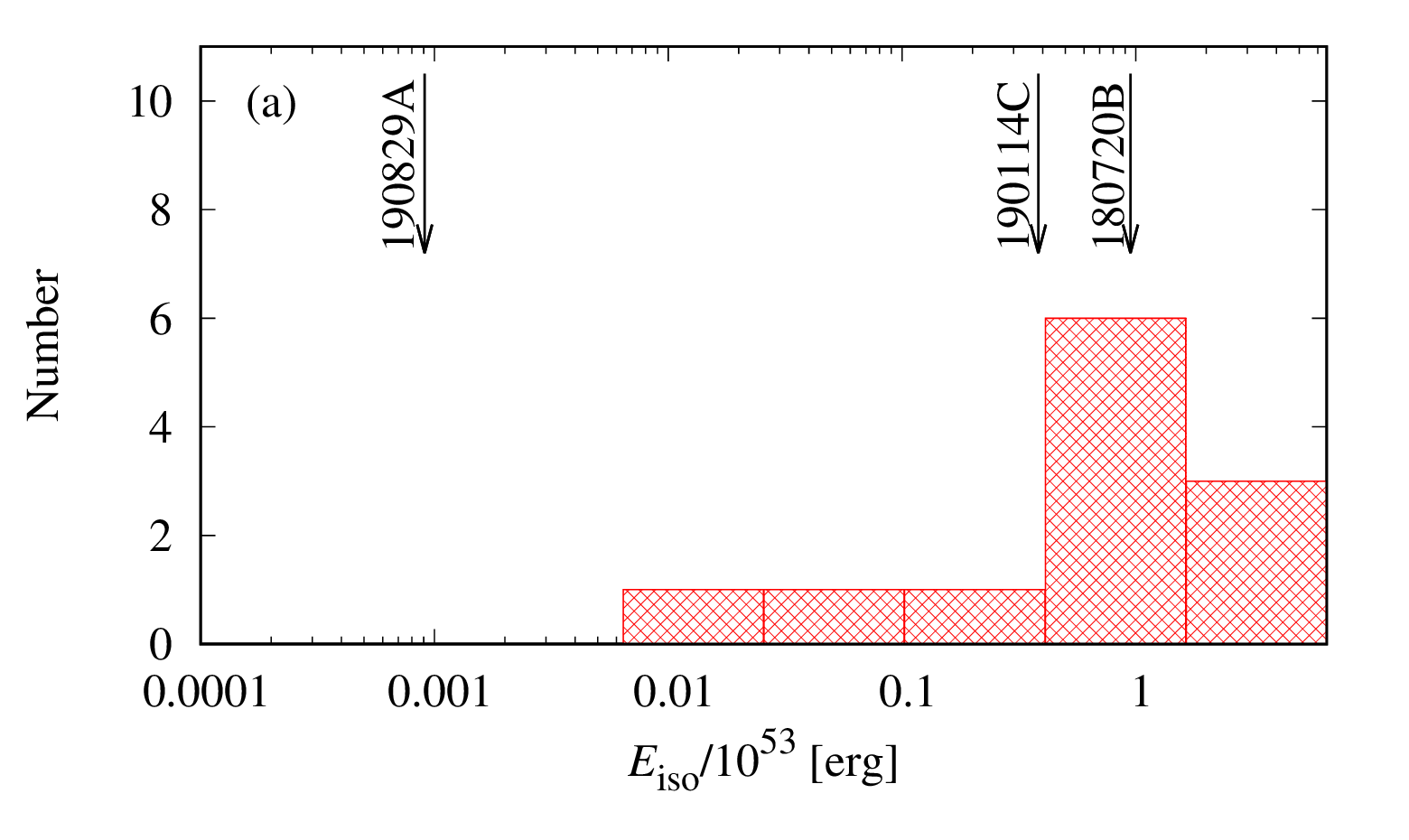}
\includegraphics[width=0.5\textwidth]{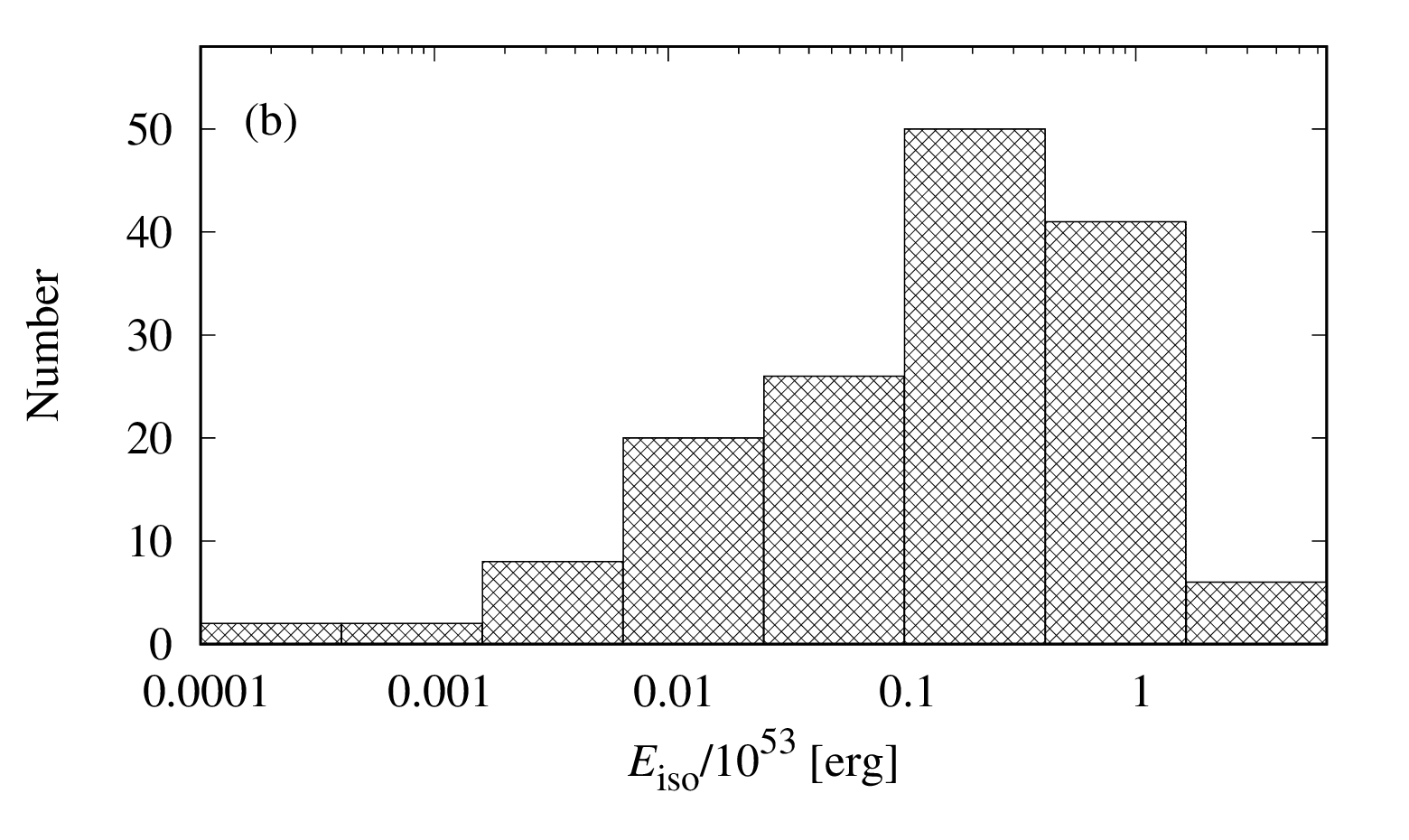}
\includegraphics[width=0.5\textwidth]{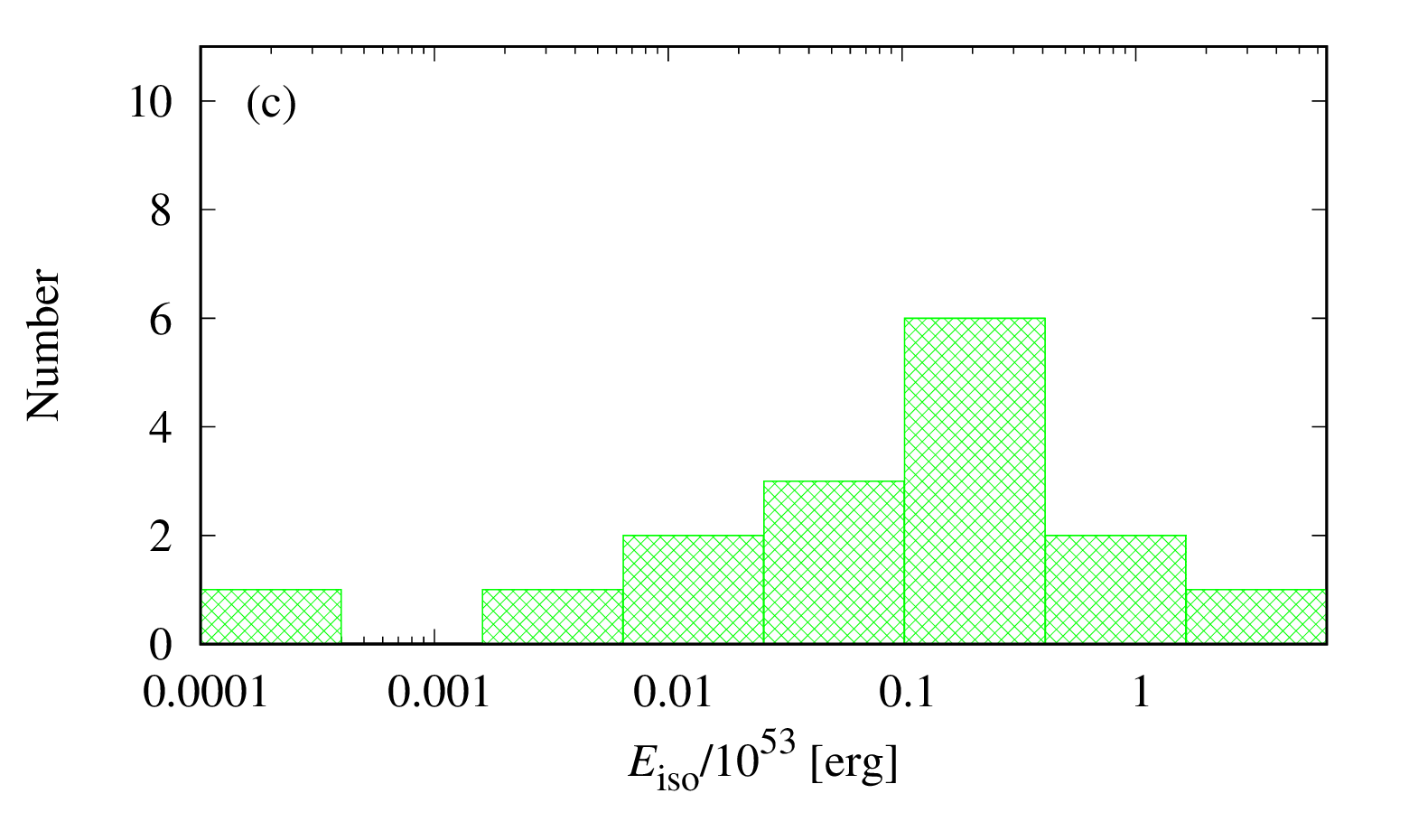}
\caption{
The distributions of isotropic gamma-ray energy of the prompt emission, $E_{\rm iso}$, for events with
measured redshift in Sample~A [panel~(a)], Sample~B [panel~(b)] and Sample~C [panel~(c)].
Arrows in panel (a) describe the values of three VHE events.
}
\label{fig:Eisodist}
\end{figure}

It would be interesting to investigate whether the difference in the X-ray behavior seen in this study is
found between more and less energetic/powerful GRBs or rather between GRBs with (very-)high-energy gamma-ray emission
and GRBs not producing the high-energy radiation.
In order to approach this issue, we plot in Figure~\ref{fig:Eisodist} the distribution of isotropic equivalent gamma-ray energy
in 15--150~keV, $E_{\rm iso}$,  of GRBs with measured redshift.
A Kolmogorov-Smirnov (KS) tests on the two distributions from different samples are performed.
For $E_{\rm iso}$ distributions of Sample~A and Sample~B (Sample~C), we get 0.33~\% (1.1~\%)
probability of being drawn from a common population.
Hence, it is indicated that Sample~A bursts tend to have larger $E_{\rm iso}$.
Another KS test on the $E_{\rm iso}$ distributions of Samples~B and C gave the probability of 79~\%,
so that we find no difference between Samples B~and C with respect to the burst energy.
Therefore, differences between GeV/TeV events (Sample~A) and general samples (Samples~B and C)
might be a result of the difference between more and less energetic GRBs.

Our present result may constrain models of the shallow decay phase of the X-ray afterglow.
In the context of the energy injection model \citep{nousek2006,zhang2006,granot2006b,kobayashi2007},
initial outflow energy is small, so that the X-ray afterglow arising from the external shock is initially dim.
If the additional energy is injected to the flow, then the X-ray afterglow becomes brighter than that in the case
of no energy injection, resulting in the shallow decay phase.
As seen in the previous paragraph,
high-energy gamma-ray events tend to have larger isotropic gamma-ray energy of the prompt emission
\citep[see also][]{ackermann2013,atteia2017,nava2014,ajello2019},
hence it is expected that the initial outflow energy is also large.
In this case, the X-ray afterglow is already bright from the beginning, and it shows no shallow decay phase.
Therefore, this model naturally explains the present result that a large fraction of events of our sample
have no clear shallow decay phase.
Some other models will be challenged if more data are accumulated in future.

%\textcolor{red}{
%We note that X-ray afterglow of GRB~190829A, one of 3 VHE events, is somewhat peculiar and complex.
%Initial small flux variation ($\la10^3$~s after the burst onset) is followed by a rising part, and  the light curve has a
%maximum at $\sim1.5$~ks after the BAT trigger.
%This observed behavior reminds us the off-axis afterglow as seen for a binary neutron star merger event, GW170817
%(???references???).
%After the maximum, the X-ray flux of GRB~190829A monotonically decreases without any light-curve break.
%Hence we fitted the light curve of this event after the maximum ($t_1=2$~ks and $t_2=100$~ks) in order to avoid the initial flux variation
%and the rising part.
%Indeed, GRB~190829A showed low  isotropic equivalent gamma-ray energy,
%$E_{\rm iso}\sim2\times10^{50}$~erg (???reference???).
%Despite such low radiation energy, this dim event could be detected because it was nearby
%(its redshift of about 0.08: ???reference???).
%Another interpretation is the shock break out emission (V. Chand et al.: arXiv:2001.00648).
%Further studies are necessary to clarify the origin of this event.
%}

We also search for any correlation between X-ray light curve parameters like $\alpha_1$ and $t_{\rm b}$ and 
GeV properties listed in the second catalog of LAT-detected GRBs  \citep{ajello2019}, such as
the temporal decay index $\alpha_{\rm GeV}$, spectral index $\beta$, and 
isotropic energy of the gamma-ray emission in the LAT energy band $E_{\rm iso}$.
Among $2~(\alpha_1~{\rm and}~t_{\rm b})\times3~(\alpha_{\rm GeV},~\beta~{\rm and}~E_{\rm iso})=6$ combinations, 
we find no statistically significant correlation  because of small sample size.
More events are necessary to have larger sample, and further analysis with better statistics is left for future work.

%%%CONCLUSION
%\section{Conclusions}

%%%%%%%%%%%%%%%%%%%%%%%%%%%%%%%%%%%%%%%%%%%%%%%%%%%%%%%%%%%%%%%%%%%%%%%%%

%%% ACKNOWLEDGEMENTS
\section*{Acknowledgments}

We would like to thank Drs.~K.~Asano, K.~Ioka, Y.~Ohira and S.~J.~Tanaka for valuable comments to improve this paper.
This work made use of data supplied by the UK Swift Science Data Centre at the University of Leicester.
 We also thank the referee for his or her thoughtful comments to improve the paper.
This work is supported in part by grant-in-aid from the Ministry of Education, Culture, Sports, Science,
and Technology (MEXT) of Japan, No.18H01232(RY), No.17H06362(TS), No.18H01257(TS), and No.17K05402(MS).
R.Y.  and T.S. deeply appreciate Aoyama Gakuin University Research Institute for helping our research by the fund.

%%%%%%%%%%%%%%%%%%%%%%%%%%%%%%%%%%%%%%%%%%%%%%%%%%%%%%%%%%%%%%%%%%%%%%%%%

\

%%% FACILITIES

{\it Facilities}: {\it Swift} and  {\it Fermi}. 

%\newpage

%%%%%%%%%%%%%%%%%%%%%%%%%%%%%%%%%%%%%%%%%%%%%%%%%%%%%%%%%%%%%%%%%%%%%%%%%

%%% BIBLIOGRAPHY

\label{lastpage}

%%%%%%%%%%%%%%%%%%%%%%%%%%%%%%%%%%%%%%%%%%%%%%%%%%%%%%%%%%%%%%%%%%%%%%%%%

\begin{table*}
 \begin{minipage}{170mm}
\label{table2a}
%\centering
  \caption{Best-fitting model parameters of GRBs in Sample~B.}
   \begin{tabular}{lcccccccc}
     \hline
    \hline
    GRB  &  $t_1$--$t_2$ [ks]   &   \multicolumn{2}{c}{SPL model : $f_{\rm S}(t)$}
    &  \multicolumn{4}{c}{DPL model: $f_{\rm D}(t)$} &   $\Delta\chi^2$~${}^{\rm a}$  \\
     \hline
     & & $\alpha_1$ & $\chi^2/dof$ & $\alpha _1$ & $\alpha _2$ & $t_b [ks]$ & $\chi^2/dof$ & \\
    \hline
    \multicolumn{9}{l}{\bf Single power-law (SPL) events}  \\
    080804 & 0.1--1000 & $1.09\pm{0.01}$ & 105/101 & $1.02\pm{0.25}$ & $1.15\pm{0.16}$ & $1.0\pm{16.8}$ & 104/99 & 1 \\
    080805${}^{\rm d}$ & 3--1000 & $0.94\pm{0.03}$ & 27/22 & $-0.08\pm{3.14}$ & $0.97\pm{0.06}$ & $4.55\pm{1.87}$ & 26/20 & 1 \\
    080916A & 0.2--1000 & $0.98\pm{0.02}$ & 170/69 & & & &  & $<1$ \\
    081118${}^{\rm e}$ & 0.9--1000 & $0.60\pm{0.05}$ & 13/8 & $0.49\pm{0.25}$ & $0.77\pm{0.75}$ & $138\pm{1210}$ & 11/6 & 2\\
    081210${}^{\rm d}$ & 5--200 & $0.67\pm{0.03}$ & 18/21 & & & & & $<1$ \\
    090417B${}^{\rm d}$ & 5--1000 & $1.38\pm{0.02}$ & 126/118 & & & & & $<1$ \\
    090423${}^{\rm d}$ & 3--500 & $1.37\pm{0.04}$ & 39/38 & $1.13\pm{0.31}$ & $1.60\pm{0.24}$ & $17.6\pm{40.8}$ & 34/36 & 5 \\
    090715B${}^{\rm d}$ & 0.45--2000 & $1.18\pm{0.02}$ & 149/74 & $1.09\pm{0.29}$ & $1.25\pm{0.19}$ & $5.5\pm{75.7}$ & 145/72 & 4 \\
    090812 & 0.4--200 & $1.23\pm{0.02}$ & 227/92 & & & & & $<1$ \\
    090814A & 0.1--200 & $2.05\pm{0.04}$ & 623/113 & & & & & $<1$ \\
    090926B & 0.5--10 & $1.17\pm{0.06}$ & 30/25 & & & & & $<1$ \\
    100513A & 0.5--1000 & $0.89\pm{0.04}$ & 91/28 & & & & & $<1$ \\
    101213A${}^{\rm e}$ & 0.5--1000 & $0.91\pm{0.02}$ & 102/71 & $0.43\pm{1.22}$ & $0.94\pm{0.05}$ & $0.65\pm{0.72}$ & 98/69 & 4 \\
    110205A & 0.8--300 & $1.63\pm{0.01}$ & 188/145 & & & & & $<1$ \\
    110801A${}^{\rm d}$ & 0.55--100 & $1.41\pm{0.02}$ & 411/111 & & & & & $<1$ \\
    111107A${}^{\rm d}$ & 0.7--100 & $0.83\pm{0.04}$ & 15/13 & $0.10\pm{0.79}$ & $0.93\pm{0.09}$ & $1.32\pm{0.91}$ & 12/11 & 3 \\
    111225A${}^{\rm d}$ & 0.7--100 & $1.40\pm{0.08}$ & 46/10 & & & & & $<1$ \\
    120712A & 0.8--500 & $1.20\pm{0.04}$ & 86/33 & $-0.56\pm{2.20}$ & $1.25\pm{0.06}$ & $1.06\pm{0.26}$ & 80/31 & 6 \\
    121201A & 0.1--100 & $1.25\pm{0.03}$ & 74/33 & & & & & $<1$ \\
    121211A${}^{\rm d}$ & 3--400 & $0.91\pm{0.04}$ & 128/68 & & & & & $<1$ \\
    130131B${}^{\rm d}$ & 0.2--20 & $1.00\pm{0.07}$ & 16/9 & $0.02\pm{0.70}$ & $1.25\pm{0.27}$ & $0.64\pm{0.54}$ & 10/7 & 6 \\
    130427B & 0.2--100 & $1.07\pm{0.04}$ & 34/19 & & & & & $<1$ \\
    130610A & 1--300 & $1.19\pm{0.03}$ & 20/26 & $-0.20\pm{2.67}$ & $1.23\pm{0.06}$ & $1.36\pm{0.46}$ & 19/24 & 1 \\
    130701A & 0.07--70 & $1.16\pm{0.01}$ & 250/128 & $1.12\pm{0.05}$ & $1.41\pm{0.30}$ & $17.9\pm{39.7}$ & 242/126 & 8 \\
    131103A${}^{\rm d}$ & 1--100 & $1.13\pm{0.04}$ & 60/36 & & & & & $<1$ \\  
    140304A${}^{\rm d}$ & 1--300 & $1.31\pm{0.12}$ & 478/24 & & & & & $<1$ \\
    140515A${}^{\rm d}$ & 6--300 & $1.20\pm{0.06}$ & 88/38 & & & & & $<1$ \\
    140907A${}^{\rm d}$ & 3--1000 & $0.98\pm{0.03}$ & 88/57 & $0.78\pm{0.98}$ & $1.07\pm{0.39}$ & $7.2\pm{71.6}$ & 87/55 & 1 \\
    141109A & 0.1--3000 & $1.40\pm{0.01}$ & 3693/497 & & & & & $<1$ \\
    141221A${}^{\rm d}$ & 3--200 & $1.17\pm{0.05}$ & 21/16 & & & & & $<1$ \\
    150301B & 0.2--80 & $1.06\pm{0.03}$ & 16/16 & $0.07\pm{1.20}$ & $1.12\pm{0.08}$ & $0.26\pm{0.15}$ & 13/14 & 3 \\
    150727A & 2--1000 & $0.81\pm{0.09}$ & 22/8 & & & & & $<1$ \\
    150818A${}^{\rm d}$ & 4--1000 & $0.73\pm{0.03}$ & 20/14 & & & & & $<1$ \\
    150821A${}^{\rm d}$ & 5--300 & $1.32\pm{0.04}$ & 31/30 & $1.14\pm{0.38}$ & $1.60\pm{0.46}$ & $33.8\pm{115}$ & 29/28 & 2 \\
    160117B & 1--300 & $0.91\pm{0.03}$ & 37/29 & & & & & $<1$ \\
    160131A & 0.05--600 & $1.18\pm{0.01}$ & 1504/596 & & & & & $<1$ \\
    160203A${}^{\rm e}$ & 4--500 & $1.16\pm{0.06}$ & 8/12 & & & & & $<1$ \\
    160425A & 0.5--1000 & $0.95\pm{0.02}$ & 130/65 & & & & & $<1$ \\
    170604A${}^{\rm d}$ & 2--2000 & $1.09\pm{0.02}$ & 149/92 & & & & & $<1$ \\
    171222A${}^{\rm e}$ & 1--1000 & $0.61\pm{0.04}$ & 56/15 & $0.57\pm{0.06}$ & $1.56\pm{1.91}$ & $540\pm{552}$ & 51/13 & 5 \\
    180205A & 0.3--1000 & $0.98\pm{0.02}$ & 72/37 & $0.65\pm{0.49}$ & $1.07\pm{0.06}$ & $1.02\pm{1.85}$ & 63/35 & 9 \\
    \hline
    \multicolumn{9}{l}{\bf Double power-law (DPL) events}  \\
    080810 & 0.4--1000 & $1.25\pm{0.04}$ & 304/71 & $0.77\pm{0.05}$ & $1.63\pm{0.05}$ & $6.02\pm{1.24}$ & 76/69 & 228 \\
    080905B & 0.2--1000 & $0.95\pm{0.03}$ & 890/85 & $0.33\pm{0.08}$ & $1.44\pm{0.04}$ & $4.84\pm{1.28}$ & 152/83 & 738\\
    080928 & 3--100 & $1.51\pm{0.06}$ & 187/72 & $0.39\pm{0.56}$ & $1.84\pm{0.11}$ & $7.41\pm{1.85}$ & 145/70 & 42 \\
    081007 & 0.7--1000 & $0.87\pm{0.03}$ & 156/50 & $0.64\pm{0.05}$ & $1.31\pm{0.10}$ & $43.0\pm{18.5}$ & 65/48 & 91 \\
    081008 & 0.5--1000 & $1.06\pm{0.04}$ & 235/53 & $0.83\pm{0.03}$ & $2.00\pm{0.10}$ & $18.8\pm{3.0}$ & 49/51 & 186 \\
    081221 & 0.3--1000 & $1.21\pm{0.01}$ & 478/247 & $0.07\pm{0.18}$ & $1.29\pm{0.02}$ & $0.53\pm{0.06}$ & 296/245 & 182 \\
    081222 & 0.1--100 & $1.03\pm{0.01}$ & 654/368 & $0.65\pm{0.08}$ & $1.13\pm{0.02}$ & $0.38\pm{0.14}$ & 492/366 & 162 \\
    090113 & 0.1--400 & $1.10\pm{0.04}$ & 100/25 & $0.13\pm{0.22}$ & $1.28\pm{0.05}$ & $0.55\pm{0.16}$ & 38 /23 & 62 \\
    090205 & 0.4--100 & $0.74\pm{0.05}$ & 125/31 & $-0.08\pm{0.19}$ & $1.19\pm{0.10}$ & $2.59\pm{0.62}$ & 50/29 & 75  \\
    090407 & 0.9--1000 & $0.65\pm{0.04}$ & 599/111 & $0.36\pm{0.03}$ & $1.63\pm{0.09}$ & $68.1\pm{8.3}$ & 181/109 & 418 \\
    090418A & 0.2--100 & $0.93\pm{0.03}$ & 249/87 & $0.37\pm{0.10}$ & $1.47\pm{0.06}$ & $2.28\pm{0.43}$ & 94/85 & 155 \\
    090424 & 0.1--5000 & $1.05\pm{0.01}$ & 1421/658 & $0.83\pm{0.02}$ & $1.21\pm{0.01}$ & $2.52\pm{0.65}$ & 828/656 & 593 \\
    090426${}^{\rm c}$ & 0.1--600 & $0.96\pm{0.04}$ & 54/28 & $-0.31\pm{0.54}$ & $1.04\pm{0.04}$ & $0.25\pm{0.06}$ & 31/26 & 23  \\
    090530 & 0.4--1000 & $0.76\pm{0.04}$ & 171/46 & $0.51\pm{0.04}$ & $1.29\pm{0.12}$ & $41.2\pm{14.5}$ & 63/44 & 108 \\
    090618 & 0.3--3000 & $1.07\pm{0.01}$ & 4158/1151 & $0.80\pm{0.01}$ & $1.47\pm{0.01}$ & $9.79\pm{0.72}$ & 1739/1149 & 2419 \\
    090809 & 10--300 & $1.22\pm{0.13}$ & 39/13 & $-0.61\pm{0.59}$ & $1.63\pm{0.15}$ & $17.2\pm{3.49}$ & 10/11 & 29 \\
    091018 & 0.03--1000 & $0.99\pm{0.01}$ & 656/139 & $0.43\pm{0.06}$ & $1.20\pm{0.02}$ & $0.61\pm{0.14}$ & 232/137 & 424 \\
    091020 & 0.3--300 & $1.07\pm{0.01}$ & 315/171 & $1.02\pm{0.02}$ & $1.71\pm{0.13}$ & $40.3\pm{9.8}$ & 240/169 & 75 \\
    091029 & 0.6--1000 & $0.73\pm{0.03}$ & 727/125 & $0.18\pm{0.05}$ & $1.15\pm{0.03}$ & $11.6\pm{1.6}$ & 177/123 & 550 \\
       \hline
    \end{tabular}
\end{minipage}
\end{table*}
\begin{table*}
 \begin{minipage}{170mm}
%\label{table2b}
%\centering
  \contcaption{}
  \begin{tabular}{lcccccccc}
  \hline 
  \hline
  GRB  &  $t_1$--$t_2$ [ks]   &   \multicolumn{2}{c}{SPL model : $f_{\rm S}(t)$}
    &  \multicolumn{4}{c}{DPL model: $f_{\rm D}(t)$} &   $\Delta\chi^2$~${}^{\rm a}$  \\
     \hline
     & & $\alpha_1$ & $\chi^2/dof$ & $\alpha _1$ & $\alpha _2$ & $t_b [ks]$ & $\chi^2/dof$ & \\
    \hline
    091109A & 0.3--1000 & $0.91\pm{0.03}$ & 29/24 & $0.36\pm{0.43}$ & $1.02\pm{0.04}$ & $1.13\pm{0.82}$ & 17/22 & 12 \\
    091208B & 0.2--1000 & $0.97\pm{0.02}$ & 199/59 & $-0.22\pm{0.25}$ & $1.15\pm{0.03}$ & $0.81\pm{0.14}$ & 86/57 & 113 \\
    100219A & 0.8--200 & $0.73\pm{0.07}$ & 184/25 & $0.56\pm{0.04}$ & $3.74\pm{0.35}$ & $33.6\pm{2.7}$ & 29/23 & 155 \\
    100302A & 1--1000 & $0.71\pm{0.04}$ & 48/23 & $0.35\pm{0.11}$ & $0.95\pm{0.07}$ & $22.3\pm{13.1}$ & 18/21 & 30 \\
    100316B & 0.1--500 & $0.92\pm{0.11}$ & 121/13 & $0.13\pm{0.14}$ & $1.34\pm{0.19}$ & $1.66\pm{1.32}$ & 18/11 & 103 \\
    100316D & 0.1--1000 & $0.63\pm{0.04}$ & 3950/402 & $0.07\pm{0.03}$ & $1.97\pm{0.04}$ & $0.83\pm{0.01}$ & 529/400 & 3421 \\
    100418A & 0.9--1000 & $0.47\pm{0.10}$ & 294/21 & $-0.18\pm{0.07}$ & $1.33\pm{0.11}$ & $71.5\pm{14.6}$ & 26/19 & 268 \\
    100424A & 0.08--30 & $1.33\pm{0.04}$ & 748/165 & $0.22\pm{0.09}$ & $2.09\pm{0.06}$ & $0.31\pm{0.02}$ & 205/163 & 543 \\ 
    100425A & 0.25--1000 & $0.69\pm{0.04}$ & 64/22 & $0.52\pm{0.05}$ & $1.19\pm{0.13}$ & $29.6\pm{14.8}$ & 21/20 & 43 \\
    100508A & 0.8--200 & $0.74\pm{0.08}$ & 516/48 & $0.34\pm{0.04}$ & $2.73\pm{0.19}$ & $22.9\pm{1.8}$ & 71/46 & 445 \\
    100615A & 0.2--200 & $3.16\pm{0.32}$ & 52920/81 & $0.32\pm{0.03}$ & $1.22\pm{0.09}$ & $15.8\pm{3.5}$ & 92/79 & 52828 \\
    100621A & 0.4--2000 & $0.95\pm{0.02}$ & 822/185 & $0.37\pm{0.04}$ & $1.25\pm{0.10}$ & $18.2\pm{4.1}$ & 85/76 & 737 \\
    100724A${}^{\rm c}$ & 0.1--400 & $1.03\pm{0.05}$ & 60/17 & $0.70\pm{0.13}$ & $1.44\pm{0.12}$ & $4.14\pm{2.53}$ & 27/15 & 33 \\
    100728B & 0.08--100 & $1.05\pm{0.02}$ & 68/37 & $0.95\pm{0.08}$ & $1.65\pm{0.34}$ & $5.83\pm{5.64}$ & 41/35 & 27 \\
    100814A & 0.5--1000 & $0.73\pm{0.02}$ & 1848/320 & $0.46\pm{0.02}$ & $2.10\pm{0.08}$ & $144\pm{8.3}$ & 524/318 & 1324 \\
    100901A & 0.01--2 & $0.96\pm{0.03}$ & 1416/233 & $-0.73\pm{0.10}$ & $1.46\pm{0.03}$ & $29.6\pm{1.21}$ & 363/231 & 1053 \\
    100905A & 0.6--100 & $0.90\pm{0.07}$ & 32/12 & $0.15\pm{0.51}$ & $1.15\pm{0.16}$ & $2.07\pm{1.38}$ & 21/10 & 11 \\
    100906A & 0.3--200 & $1.07\pm{0.04}$ & 1055/132 & $0.71\pm{0.04}$ & $2.02\pm{0.09}$ & $12.7\pm{1.6}$ & 348/130 & 707 \\
    101219B & 0.1--10 & $1.62\pm{0.06}$ & 1073/141 & $0.30\pm{0.17}$ & $2.41\pm{0.08}$ & $0.33\pm{0.02}$ & 452/139 & 621 \\ 
    110106B & 0.2--1000 & $0.94\pm{0.03}$ & 228/53 & $0.59\pm{0.06}$ & $1.42\pm{0.08}$ & $13.5\pm{4.4}$ & 77/51 & 151 \\
    110503A & 0.1--1000 & $1.12\pm{0.01}$ & 612/392 & $1.05\pm{0.01}$ & $1.48\pm{0.05}$ & $34.4\pm{12.2}$ & 487/390 & 125 \\
    110715A & 0.08--1000 & $0.91\pm{0.01}$ & 799/256 & $0.21\pm{0.15}$ & $0.97\pm{0.01}$ & $0.22\pm{0.05}$ & 606/254 & 193 \\
    110808A & 0.4--1000 & $0.67\pm{0.04}$ & 24/12 & $0.57\pm{0.04}$ & $1.41\pm{0.33}$ & $167\pm{79.1}$ & 9/10 & 15 \\
    110818A${}^{\rm b}$ & 0.5--200 & $1.18\pm{0.03}$ & 112/59 & $1.15\pm{0.03}$ & $2.37\pm{0.65}$ & $66.9\pm{29.7}$ & 99/57 & 13 \\
    111008A & 0.3--1000 & $0.88\pm{0.03}$ & 928/136 & $0.29\pm{0.05}$ & $1.30\pm{0.03}$ & $7.42\pm{1.10}$ & 203/134 & 725 \\
    111123A & 4--200 & $1.08\pm{0.05}$ & 124/42 & $0.73\pm{0.09}$ & $1.93\pm{0.21}$ & $34.9\pm{10.1}$ & 56/40 & 68 \\
    111129A & 0.1--350 & $0.87\pm{0.03}$ & 212/46 & $0.47\pm{0.12}$ & $1.23\pm{0.05}$ & $2.93\pm{1.43}$ & 65/44 & 147 \\
    111228A & 0.4--1000 & $0.88\pm{0.02}$ & 767/148 & $0.39\pm{0.03}$ & $1.25\pm{0.03}$ & $12.9\pm{1.8}$ & 179/146 & 588 \\
    111229A & 0.2--20 & $0.34\pm{0.11}$ & 314/35 & $-0.13\pm{0.08}$ & $2.15\pm{0.40}$ & $6.94\pm{1.05}$ & 74/33 & 240 \\
    120118B & 0.3--100 & $0.53\pm{0.05}$ & 160/35 & $-0.50\pm{0.17}$ & $1.02\pm{0.07}$ & $2.31\pm{0.34}$ & 41/33 & 119 \\
    120119A${}^{\rm b}$ & 0.15--100 & $1.01\pm{0.01}$ & 201/91 & $1.00\pm{0.02}$ & $2.46\pm{0.65}$ & $32.4\pm{8.1}$ & 180/89 & 21 \\
    120211A & 1--100 & $0.63\pm{0.16}$ & 92/12 & $-0.25\pm{0.19}$ & $1.29\pm{0.17}$ & $7.71\pm{2.07}$ & 17/10 & 75 \\
    120224A & 0.3--600 & $0.73\pm{0.03}$ & 390/79 & $-0.59\pm{0.16}$ & $0.98\pm{0.03}$ & $1.94\pm{0.23}$ & 115/77 & 275 \\
    120326A & 3--1000 & $0.25\pm{0.05}$ & 1717/193 & $-0.24\pm{0.03}$ & $2.04\pm{0.07}$ & $51.7\pm{1.9}$ & 250/191 & 1467 \\
    120327A & 0.2--200 & $1.02\pm{0.03}$ & 259/59 & $0.65\pm{0.08}$ & $1.53\pm{0.07}$ & $3.36\pm{0.92}$ & 105/57 & 154 \\
    120404A & 0.4--200 & $0.94\pm{0.07}$ & 175/37 & $0.24\pm{0.18}$ & $1.81\pm{0.10}$ & $3.17\pm{0.61}$ & 45/35 & 130 \\
    120422A & 1--2000 & $0.63\pm{0.08}$ & 35/9 & $0.22\pm{0.08}$ & $1.21\pm{0.14}$ & $153\pm{54}$ & 4/7 & 31 \\
    120521C & 0.7--100 & $0.62\pm{0.12}$ & 37/8 & $0.34\pm{0.07}$ & $2.68\pm{0.64}$ & $22.1\pm{3.8}$ & 4/6 & 33 \\
    120811C & 0.2--100 & $0.87\pm{0.02}$ & 100/50 & $0.54\pm{0.09}$ & $1.16\pm{0.10}$ & $3.15\pm{1.74}$ & 66/48 & 34 \\
    120907A & 0.1--300 & $0.89\pm{0.02}$ & 244/88 & $0.50\pm{0.10}$ & $1.10\pm{0.04}$ & $2.01\pm{0.87}$ & 140/86 & 104 \\
    121024A & 3--300 & $1.13\pm{0.05}$ & 88/46 & $0.83\pm{0.14}$ & $1.58\pm{0.24}$ & $30.2\pm{30.2}$ & 63/44 & 25 \\
    121027A & 30--3000 & $1.06\pm{0.04}$ & 246/79 & $0.45\pm{0.13}$ & $1.56\pm{0.08}$ & $166\pm{39}$ & 101/77 & 145 \\
    121128A & 0.15--100 & $1.13\pm{0.04}$ & 656/93 & $0.50\pm{0.05}$ & $1.62\pm{0.04}$ & $1.46\pm{0.20}$ & 115/91 & 541 \\
    121209A & 0.1--60 & $0.82\pm{0.03}$ & 383/88 & $-0.67\pm{0.19}$ & $1.22\pm{0.03}$ & $0.87\pm{0.11}$ & 99/86 & 284 \\
    130408A & 0.1--100 & $0.87\pm{0.05}$ & 718/57 & $0.28\pm{0.33}$ & $1.46\pm{0.11}$ & $2.51\pm{2.16}$ & 301/55 & 417 \\
    130418A & 0.1--500 & $1.16\pm{0.02}$ & 23/96 & $0.72\pm{0.10}$ & $1.41\pm{0.09}$ & $0.68\pm{0.34}$ & 166/94 & 67 \\
    130505A & 2--2000 & $1.37\pm{0.01}$ & 557/272 & $0.92\pm{0.05}$ & $1.61\pm{0.03}$ & $22.4\pm{4.2}$ & 288/270 & 269 \\
    130511A & 0.07--40 & $1.02\pm{0.05}$ & 65/26 & $-0.70\pm{0.59}$ & $1.13\pm{0.05}$ & $0.18\pm{0.03}$ & 31/24 & 34 \\
    130603B${}^{\rm c}$ & 0.04--200 & $0.78\pm{0.03}$ & 524/76 & $0.31\pm{0.07}$ & $1.52\pm{0.07}$ & $2.28\pm{0.36}$ & 155/74 & 69 \\
    130606A & 4--300 & $2.40\pm{0.20}$ & 1017/37 & $-0.23\pm{0.68}$ & $1.76\pm{0.09}$ & $9.12\pm{1.79}$ & 38/35 & 979 \\
    130612A & 0.7--50 & $0.92\pm{0.05}$ & 13/11 & $0.40\pm{0.16}$ & $1.26\pm{0.08}$ & $3.32\pm{1.14}$ & 2/9 & 11 \\
    130925A & 20--10000 & $1.02\pm{0.01}$ & 923/441 & $0.75\pm{0.04}$ & $1.40\pm{0.04}$ & $322\pm{66}$ & 612/439 & 311 \\
    131030A & 0.4--2000 & $1.10\pm{0.01}$ & 461/264 & $0.78\pm{0.06}$ & $1.25\pm{0.02}$ & $2.60\pm{0.90}$ & 344/262 & 117 \\
    131117A & 0.2--200 & $0.93\pm{0.03}$ & 21/14 & $0.05\pm{0.39}$ & $1.03\pm{0.05}$ & $0.57\pm{0.23}$ & 10/12 & 11 \\
    140114A & 1--300 & $4.62\pm{0.76}$ & 16053/32 & $0.17\pm{0.10}$ & $1.21\pm{0.13}$ & $18.8\pm{5.8}$ & 41/30 & 16012 \\
    140206A & 0.5--2000 & $0.99\pm{0.01}$ & 1598/476 & $0.56\pm{0.05}$ & $1.29\pm{0.02}$ & $4.60\pm{0.74}$ & 751/474 & 846 \\
    140318A & 0.1--100 & $1.56\pm{0.05}$ & 256/54 & $0.20\pm{0.34}$ & $1.77\pm{0.06}$ & $0.24\pm{0.05}$ & 109/52 & 147 \\
    140419A & 0.8--2000 & $1.14\pm{0.01}$ & 642/253 & $0.72\pm{0.07}$ & $1.40\pm{0.03}$ & $5.47\pm{1.26}$ & 340/251 & 302 \\
    140430A & 3--200 & $0.75\pm{0.04}$ & 57/39 & $0.60\pm{0.10}$ & $1.53\pm{0.43}$ & $47.7\pm{22.1}$ & 45/37 & 12 \\
    140506A & 0.7--3000 & $0.91\pm{0.01}$ & 246/162 & $0.51\pm{0.30}$ & $0.95\pm{0.03}$ & $1.62\pm{1.22}$ & 229/160 & 17 \\
       \hline
     \end{tabular}
\end{minipage}
\end{table*}
\begin{table*}
 \begin{minipage}{170mm}
%\label{table2c}
%\centering
  \contcaption{}
  \begin{tabular}{lcccccccc}
  \hline 
  \hline
  GRB  &  $t_1$--$t_2$ [ks]   &   \multicolumn{2}{c}{SPL model : $f_{\rm S}(t)$}
    &  \multicolumn{4}{c}{DPL model: $f_{\rm D}(t)$} &   $\Delta\chi^2$~${}^{\rm a}$  \\
     \hline
     & & $\alpha_1$ & $\chi^2/dof$ & $\alpha _1$ & $\alpha _2$ & $t_b [ks]$ & $\chi^2/dof$ & \\
    \hline
    140512A & 0.3--300 & $0.92\pm{0.01}$ & 1206/377 & $0.73\pm{0.02}$ & $1.57\pm{0.05}$ & $13.2\pm{1.6}$ & 482/375 & 724 \\
    140518A & 0.2--20 & $0.79\pm{0.05}$ & 138/37 & $0.36\pm{0.08}$ & $1.63\pm{0.22}$ & $3.20\pm{0.72}$ & 57/35 & 81 \\
    140614A & 5--200 & $1.32\pm{0.09}$ & 53/23 & $-0.71\pm{1.18}$ & $1.60\pm{0.13}$ & $7.42\pm{1.28}$ & 35/21 & 18 \\
    140629A & 0.1--100 & $0.98\pm{0.02}$ & 466/112 & $0.81\pm{0.02}$ & $1.79\pm{0.08}$ & $7.72\pm{1.26}$ & 140/110 & 326 \\
    140703A & 2--100 & $1.38\pm{0.06}$ & 308/73 & $0.57\pm{0.14}$ & $2.13\pm{0.13}$ & $13.6\pm{2.2}$ & 120/71 & 188 \\
    141004A & 0.04--100 & $1.02\pm{0.07}$ & 148/24 & $0.67\pm{0.05}$ & $2.02\pm{0.17}$ & $3.59\pm{0.68}$ & 25/22 & 123 \\
    141121A & 100--1000 & $1.22\pm{0.17}$ & 36/12 & $0.02\pm{0.35}$ & $2.61\pm{0.35}$ & $318\pm{46}$ & 8/10 & 28 \\
    141220A & 0.08--40 & $1.16\pm{0.03}$ & 58/28 & $-0.29\pm{0.26}$ & $1.40\pm{0.04}$ & $0.20\pm{0.03}$ & 16/26 & 42 \\
    150323A & 0.35--400 & $0.77\pm{0.06}$ & 99/21 & $0.52\pm{0.08}$ & $1.29\pm{0.20}$ & $17.5\pm{12.5}$ & 44/19 & 55 \\
    150910A & 0.2--300 & $0.95\pm{0.02}$ & 2898/352 & $0.43\pm{0.03}$ & $2.25\pm{0.05}$ & $6.17\pm{0.35}$ & 694/350 & 2204 \\
    151021A & 0.2--300 & $1.15\pm{0.01}$ & 279/118 & $0.91\pm{0.07}$ & $1.42\pm{0.05}$ & $2.67\pm{1.48}$ & 168/116 & 111 \\
    151027A & 0.5--1000 & $0.93\pm{0.02}$ & 3546/453 & $0.06\pm{0.03}$ & $1.65\pm{0.02}$ & $3.92\pm{0.15}$ & 642/451 & 2904 \\
    151027B & 2--500 & $0.89\pm{0.04}$ & 87/44 & $0.67\pm{0.07}$ & $1.46\pm{0.18}$ & $49.2\pm{23.3}$ & 50/42 & 37 \\
    151029A & 0.1--30 & $1.09\pm{0.08}$ & 32/8 & $0.61\pm{0.25}$ & $1.45\pm{0.31}$ & $1.22\pm{1.38}$ & 15/6 & 17 \\
    151215A & 0.4--200 & $0.99\pm{0.06}$ & 23/12 & $-0.09\pm{0.47}$ & $1.13\pm{0.06}$ & $1.02\pm{0.34}$ & 8/10 & 15 \\
    160121A & 0.2--500 & $0.41\pm{0.04}$ & 53/24 & $0.29\pm{0.04}$ & $1.84\pm{0.50}$ & $18.9\pm{4.2}$ & 22/22 & 31 \\
    160227A & 1.5--1000 & $0.76\pm{0.03}$ & 437/103 & $0.19\pm{0.08}$ & $1.11\pm{0.04}$ & $18.0\pm{3.2}$ & 172/101 & 265 \\
    160804A & 3--1000 & $0.74\pm{0.04}$ & 115/45 & $-0.27\pm{0.44}$ & $0.92\pm{0.07}$ & $9.10\pm{2.82}$ & 76/43 & 39 \\
    161014A & 0.1--100 & $0.96\pm{0.03}$ & 293/63 & $0.60\pm{0.08}$ & $1.82\pm{0.09}$ & $2.77\pm{0.58}$ & 92/61 & 201 \\
    161017A & 3--1000 & $1.23\pm{0.03}$ & 204/83 & $0.86\pm{0.09}$ & $1.75\pm{0.10}$ & $27.5\pm{7.8}$ & 110/81 & 94 \\
    161108A & 0.6--800 & $0.55\pm{0.04}$ & 81/26 & $0.34\pm{0.07}$ & $1.00\pm{0.20}$ & $63.6\pm{46.8}$ & 44/24 & 37 \\
    161117A & 0.4--2000 & $0.88\pm{0.02}$ & 495/147 & $0.67\pm{0.04}$ & $1.24\pm{0.05}$ & $16.4\pm{5.6}$ & 258/145 & 237 \\
    161129A & 0.2--30 & $0.86\pm{0.06}$ & 555/58 & $-0.09\pm{0.24}$ & $2.14\pm{0.09}$ & $2.30\pm{0.46}$ & 100/56 & 455 \\
    161219B & 0.8--20000 & $0.81\pm{0.01}$ & 1094/396 & $0.53\pm{0.03}$ & $0.97\pm{0.02}$ & $35.5\pm{9.7}$ & 640/394 & 454 \\
    170113A & 0.2--1000 & $0.86\pm{0.02}$ & 674/161 & $0.51\pm{0.04}$ & $1.26\pm{0.03}$ & $4.77\pm{0.88}$ & 227/159 & 447 \\
    170202A & 0.3--400 & $0.86\pm{0.04}$ & 191/40 & $-0.18\pm{0.14}$ & $1.15\pm{0.05}$ & $2.00\pm{0.32}$ & 50/38 & 141 \\
    170519A & 0.6--300 & $0.96\pm{0.04}$ & 309/81 & $0.40\pm{0.09}$ & $1.36\pm{0.08}$ & $8.30\pm{1.71}$ & 154/79 & 155 \\
    170531B & 1--100 & $0.78\pm{0.06}$ & 25/12 & $0.43\pm{0.46}$ & $1.09\pm{0.30}$ & $5.88\pm{12.5}$ & 14/10 & 11 \\
    170607A & 0.7--1000 & $0.68\pm{0.02}$ & 578/184 & $0.42\pm{0.03}$ & $1.07\pm{0.04}$ & $26.8\pm{6.0}$ & 237/182 & 341 \\
    170705A & 0.7--3000 & $0.85\pm{0.02}$ & 870/223 & $0.53\pm{0.04}$ & $1.20\pm{0.04}$ & $29.8\pm{6.5}$ & 421/221 & 449 \\
    170714A & 0.3--1000 & $0.91\pm{0.01}$ & 18200/1398 & $0.53\pm{0.02}$  & $3.76\pm{0.09}$ & $11.9\pm{0.3}$ & 7908/1396 & 10292 \\
    171205A & 5--3000 & $0.65\pm{0.06}$ & 369/44 & $-0.27\pm{0.11}$ & $1.10\pm{0.07}$ & $91.4\pm{15.1}$ & 87/42 & 282 \\
    180325A & 0.2--100 & $1.07\pm{0.03}$ & 999/114 & $-0.01\pm{0.17}$ & $2.05\pm{0.04}$ & $1.81\pm{0.28}$ & 138/112 & 861 \\
    180620B & 0.5--1000 & $0.73\pm{0.02}$ & 392/98 & $0.47\pm{0.04}$ & $1.21\pm{0.08}$ & $48.9\pm{12.6}$ & 180/96 & 212 \\
    180624A & 3--300 & $0.95\pm{0.07}$ & 76/34 & $0.51\pm{0.23}$ & $1.42\pm{0.19}$ & $15.5\pm{8.6}$ & 51/32 & 25\\
       \hline
  \end{tabular}
\begin{tablenotes}
\item {\bf Notes.} \\
${}^{\rm a}$~A DPL model is statistically preferred at $>3\sigma$ over a simpler SPL model when $\Delta\chi^2>10$.\\
${}^{\rm b}$~Best-fitting values of $\alpha_1$ and $\alpha_2$ of DPL model are consistent with the jet break (see section~5).\\
${}^{\rm c}$~Short GRBs.\\
${}^{\rm d}$~Events showing only X-ray flares until 3~ks after the burst trigger.\\
${}^{\rm e}$~Events showing long initial steep decay phase lasting 2~ks.\\
\end{tablenotes}
\end{minipage}
\end{table*}

%%%%%%%%%%%%%%%%%%%%%%%%%%%%%%%%%%%%%%%%%%%%%%%%%%%%%%%%%%%%%%%%%%%%%%%%%

\newpage

\begin{table*}
 \begin{minipage}{170mm}
%\centering
  \caption{Best-fitting model parameters of GRBs in Sample~C}
 \begin{tabular}{lcccccccc}
    \hline
    \hline
    GRB  &  $t_1$--$t_2$ [ks]   &   \multicolumn{2}{c}{SPL model : $f_{\rm S}(t)$}
    &  \multicolumn{4}{c}{DPL model: $f_{\rm D}(t)$} &  $\Delta\chi^2$~${}^{\rm a}$  \\
     \hline
     & & $\alpha_1$ & $\chi^2/dof$ & $\alpha _1$ & $\alpha _2$ & $t_b [ks]$ & $\chi^2/dof$ & \\
    \hline
    \multicolumn{9}{l}{\bf Single power-law (SPL) events}  \\
    081118$^{\rm e}$ & 0.9--1000 & $0.60\pm{0.05}$ & 13/8 & $0.49\pm{0.25}$ & $0.77\pm{0.75}$ & $138\pm{1210}$ & 11/6 & 2 \\
    090422 & 0.5--100 & $0.94\pm{0.03}$ & 40/28 & $-0.05\pm{0.94}$ & $1.01\pm{0.05}$ & $0.82\pm{0.34}$ & 35/26 & 5 \\
    090831C$^{\rm d}$ & 4--150 & $0.84\pm{0.06}$ & 7/9 & $-0.55\pm{3.11}$ & $0.91\pm{0.10}$ & $5.49\pm{2.23}$ & 6/7 & 1 \\	
    091221$^{\rm d}$ & 0.2--200 & $1.09\pm{0.04}$ & 10/10 & $0.99\pm{0.61}$ & $1.20\pm{0.72}$ & $8.3\pm{207}$ & 9/8 & 1 \\	
    110801A$^{\rm d}$ & 0.55--100 & $1.41\pm{0.02}$ & 411/111 & & & & & $<1$ \\
    \hline
    \multicolumn{9}{l}{\bf Double power-law (DPL) events}  \\
    080928 & 3--100 & $1.51\pm{0.06}$ & 187/72 & $0.39\pm{0.56}$ & $1.84\pm{0.11}$ & $7.41\pm{1.85}$ & 145/70 & 42 \\
    081008 & 0.5--1000 &	$1.06\pm{0.04}$ & 235/53 & $0.83\pm{0.03}$ & $2.00\pm{0.10}$ &	$18.8\pm{3.0}$ & 49/51 & 186 \\
    081126 & 0.2--800 & $0.98\pm{0.04}$ & 567/74 & $0.45\pm{0.07}$ & $1.49\pm{0.05}$ & $5.11\pm{1.26}$ & 133/72 & 434 \\	
	081127 & 0.4--50 & $0.64\pm{0.11}$ & 35/9 & $-0.53\pm{0.50}$ & $1.03\pm{0.22}$ & $2.05\pm{0.94}$ & 14/7 & 11 \\	
    081222 & 0.1--100 & $1.03\pm{0.01}$ & 654/368 & $0.65\pm{0.08}$ &	$1.13\pm{0.02}$ & $0.38\pm{0.14}$ & 492/366 & 162 \\
    090113 & 0.1--400 & $1.10\pm{0.04}$ & 100/25 & $0.13\pm{0.22}$ & $1.28\pm{0.05}$ & $0.55\pm{0.16}$ & 38 /23 & 62 \\
    090407 & 0.9--1000 &	$0.65\pm{0.04}$ & 599/111 & $0.36\pm{0.03}$ & $1.63\pm{0.09}$ & $68.1\pm{8.3}$ & 181/109 & 418 \\
    090516 & 3--200 & $1.28\pm{0.03}$ & 378/133 & $0.65\pm{0.10}$ & $1.88\pm{0.09}$ & $16.5\pm{2.6}$ & 184/131 & 194 \\	
    090518 & 0.3--100 & $0.75\pm{0.03}$ & 72/35 & $0.27\pm{0.18}$ & $1.05\pm{0.09}$ & $2.41\pm{1.16}$ & 42/33 & 30 \\	
	090529 & 2--1000 & $0.68\pm{0.06}$ & 18/8 & $0.51\pm{0.05}$ & $1.68\pm{0.31}$ & $223\pm{71}$ & 3/6 & 15 \\
	090621A & 0.8--1000 & $0.92\pm{0.03}$ & 152/53 & $0.55\pm{0.07}$ & $1.27\pm{0.06}$ & $10.6\pm{3.4}$ & 56/51 & 96 \\	
	090728 & 0.3--100 & $1.10\pm{0.06}$ & 120/26 & $0.00\pm{0.18}$ & $1.80\pm{0.09}$ & $1.81\pm{0.25}$ & 23/24 & 97 \\	
	090813 & 0.05--1000 & $1.03\pm{0.01}$ & 1666/295 & $0.23\pm{0.04}$ & $1.25\pm{0.01}$ & $0.55\pm{0.04}$ & 400/293 & 1266 \\	
    091208B & 0.2--1000 & $0.97\pm{0.02}$ & 199/59 & $-0.22\pm{0.25}$ & $1.15\pm{0.03}$ & $0.81\pm{0.14}$ & 86/57 & 113 \\
    100111A & 0.1--300 & $0.84\pm{0.03}$ & 69/33 & $0.39\pm{0.12}$ & $1.04\pm{0.06}$ & $1.38\pm{0.60}$ & 35/31 & 34 \\
    100212A & 1--100 & $0.90\pm{0.10}$ & 50/9 & $0.36\pm{0.43}$ & $1.35\pm{0.29}$ & $6.43\pm{7.81}$ & 20/7 & 30 \\	
    100316D & 0.1--1000 & $0.63\pm{0.04}$ & 3950/402 & $0.07\pm{0.03}$ & $1.97\pm{0.04}$ & $0.83\pm{0.01}$ & 529/400 & 3421 \\
    100418A & 0.9--1000 & $0.47\pm{0.10}$ & 294/21 & $-0.18\pm{0.07}$ & $1.33\pm{0.11}$ &	$71.5\pm{14.6}$ & 26/19 & 268 \\
    100614A & 1.5--1000 & $0.69\pm{0.05}$ & 130/30 & $0.56\pm{0.05}$ & $2.48\pm{0.54}$ & $202\pm{41}$ & 65/28 & 65 \\
    100704A & 0.5--2000 & $0.92\pm{0.02}$ & 450/157 & $0.67\pm{0.03}$ & $1.32\pm{0.05}$ & $33.8\pm{8.0}$ & 199/155 & 251 \\
	100725B & 0.5--500 & $0.87\pm{0.04}$ & 216/53 & $0.41\pm{0.05}$ & $1.41\pm{0.07}$ & $16.0\pm{2.6}$ & 54/51 & 162 \\
	100728B & 0.07--100 & $1.04\pm{0.02}$ & 74/38 & $0.94\pm{0.08}$ & $1.66\pm{0.35}$ & $6.04\pm{5.48}$ & 45/36 & 29 \\	
	100802A & 5--1000 & $0.55\pm{0.05}$ & 65/22 & $-0.40\pm{0.30}$ & $0.72\pm{0.07}$ & $23.2\pm{9.3}$ & 28/20 & 37 \\
	100902A &1.5--2000 & $0.77\pm{0.02}$ & 144/56 & $0.72\pm{0.02}$ & $4.44\pm{1.11}$ & $823\pm{90}$ & 95/54 & 49 \\	
	101219B & 0.1--10 & $1.62\pm{0.06}$ & 1073/141 & $0.30\pm{0.17}$ & $2.41\pm{0.08}$ & $0.33\pm{0.02}$ & 452/139 &	621 \\ 
110102A & 0.5--1000 & $0.94\pm{0.02}$ & 1290/281 & $0.49\pm{0.03}$ & $1.41\pm{0.03}$ & $13.4\pm{1.3}$ & 335/279 & 955 \\
110223A & 0.5--800 & $0.60\pm{0.03}$ & 27/22 & $0.41\pm{0.10}$ & $0.79\pm{0.10}$ & $23.5\pm{33.0}$ & 16/20 & 11 \\
	110411A & 0.3--200 & $0.86\pm{0.04}$ & 116/37 & $0.40\pm{0.11}$ & $1.24\pm{0.10}$ & $3.26\pm{1.18}$ & 56/35 & 60 \\
	110414A & 0.6--200 & $1.18\pm{0.07}$ & 138/30 & $0.05\pm{0.27}$ & $1.65\pm{0.08}$ & $2.58\pm{0.46}$ & 46/28 & 92 \\
	110808A & 0.4--1000 & $0.67\pm{0.04}$ & 24/12 & $0.57\pm{0.04}$ & $1.41\pm{0.33}$ & $167\pm{79}$ & 9/10 & 15 \\
	120118B & 0.3--100 & $0.53\pm{0.05}$ & 160/35 & $-0.50\pm{0.17}$ & $1.02\pm{0.07}$ & $2.31\pm{0.34}$ & 41/33 & 119 \\
    \hline
  \end{tabular}
\begin{tablenotes}
\item {\bf Notes.} \\
${}^{\rm a}$~A DPL model is statistically preferred at $>3\sigma$ over a simpler SPL model when $\Delta\chi^2>10$.\\
${}^{\rm b}$~Best-fitting values of $\alpha_1$ and $\alpha_2$ of DPL model are consistent with the jet break (see section~5).\\
%%${}^{\rm c}$~Short GRBs.\\
${}^{\rm d}$~Events showing only X-ray flares until 3~ks after the burst trigger.\\
${}^{\rm e}$~Events showing long initial steep decay phase lasting 2~ks.\\
\end{tablenotes}
\label{table3}
\end{minipage}
\end{table*}

%%%%%%%%%%%%%%%%%%%%%%%%%%%%%%%%%%%%%%%%%%%%%%%%%%%%%%%%%%%%%%%%%%%%%%%%%

%\begin{figure}
%%\centering \vspace*{1pt}
%\includegraphics[width=0.5\textwidth]{a1-1.eps}
%\includegraphics[width=0.5\textwidth]{a1-2.eps}
%\caption{
%TEST!!!!
%}
%\label{fig:test}
%\end{figure}

%\begin{figure}
%\includegraphics[width=85mm, bb= 0 0 700 300]{fig_test.png}
%\caption{
%TEST!!!!
%}
%\label{fig:test_png}
%\end{figure}

%\onecolumn

%\twocolumn

%%% END OF DOCUMENT
\end{document}